\documentclass{jfm}

\usepackage{graphicx}
\usepackage{natbib}

\usepackage{color,floatflt}
\usepackage{booktabs,subfigure}
\usepackage{float,amssymb,amsmath}
\usepackage{verbatim}
\usepackage{comment}
\usepackage{tikz}
\usepackage{tabularx}
\usepackage{bigstrut}
\usepackage{enumitem}

\usepackage{longtable}
\usepackage{comment}
\usepackage{soul}

\usepackage[bordercolor=black,linecolor=black,backgroundcolor=white,textwidth=3.5cm]{todonotes}
\usepackage{epstopdf}
\usepackage{lscape}

\makeatletter
\tikzset{
  RL/.style={% without calc
    rounded corners={1pt},
    to path={% we asume that you use this path only on nodes (otherwise it will get tricky)
      \pgfextra
        \pgf@process{\pgfpointanchor{\tikztostart}{east}}%
        \pgf@xa\pgf@x\pgf@ya\pgf@y
        \pgf@process{\pgfpointanchor{\tikztotarget}{east}}%
        \pgf@xb\pgf@x\pgf@yb\pgf@y
        \ifdim\pgf@xb>\pgf@xa
          \pgf@xa\pgf@xb
        \fi
        \pgfmathsetlength\pgf@xc{#1}%
        \advance\pgf@xa\pgf@xc
      \endpgfextra
      -- (+\pgf@xa,+\pgf@ya) -- (+\pgf@xa,+\pgf@yb) \tikztonodes -- (\tikztotarget)
    }
  },
  RL/.default=2cm}

\tikzset{
  LL/.style={% without calc
    rounded corners={1pt},
    to path={% we asume that you use this path only on nodes (otherwise it will get tricky)
      \pgfextra
        \pgf@process{\pgfpointanchor{\tikztostart}{west}}%
        \pgf@xa\pgf@x\pgf@ya\pgf@y
        \pgf@process{\pgfpointanchor{\tikztotarget}{west}}%
        \pgf@xb\pgf@x\pgf@yb\pgf@y
        \ifdim\pgf@xb>\pgf@xa
          \pgf@xa\pgf@xb
        \fi
        \pgfmathsetlength\pgf@xc{#1}%
        \advance\pgf@xa\pgf@xc
      \endpgfextra
      -- (+\pgf@xa,+\pgf@ya) -- (+\pgf@xa,+\pgf@yb) \tikztonodes -- (\tikztotarget)
    }
  },
  LL/.default=-2cm}

\ifCUPmtlplainloaded \else
  \checkfont{eurm10}
  \iffontfound
    \IfFileExists{upmath.sty}
      {\typeout{^^JFound AMS Euler Roman fonts on the system,
                   using the 'upmath' package.^^J}%
       \usepackage{upmath}}
      {\typeout{^^JFound AMS Euler Roman fonts on the system, but you
                   dont seem to have the}%
       \typeout{'upmath' package installed. JFM.cls can take advantage
                 of these fonts,^^Jif you use 'upmath' package.^^J}%
      }
  \else
  \fi
\fi

\ifCUPmtlplainloaded \else
  \checkfont{msam10}
  \iffontfound
    \IfFileExists{amssymb.sty}
      {\typeout{^^JFound AMS Symbol fonts on the system, using the
                'amssymb' package.^^J}%
       \usepackage{amssymb}%
       \let\le=\leqslant  \let\leq=\leqslant
       \let\ge=\geqslant  \let\geq=\geqslant
      }{}
  \fi
\fi

\ifCUPmtlplainloaded \else
  \IfFileExists{amsbsy.sty}
    {\typeout{^^JFound the 'amsbsy' package on the system, using it.^^J}%
     \usepackage{amsbsy}}
    {\providecommand\boldsymbol[1]{\mbox{\boldmath $##1$}}}
\fi

\newsavebox{\astrutbox}
\sbox{\astrutbox}{\rule[-5pt]{0pt}{20pt}}

\def\bfg#1{\setbox0=\hbox{$#1$}%----  poor man's boldface: make a letter plumpy
        \kern-.025em\copy0\kern-\wd0
        \kern.05em\copy0\kern-\wd0
        \kern-.025em\raise.0433em\box0}

\def\drwln#1#2{\raise 2.5pt\vbox{\hrule width #1pt height #2pt}}
\def\solid{\ \drwln{24}{1.0}\ }
\def\spc#1{\hskip #1pt}
\def\dashed{\ \hbox {\drwln{4}{1.0}\spc{2}
                   \drwln{4}{1.0}\spc{2}\drwln{4}{1.0}}\nobreak\ }
\def\dashdot{\ \hbox {\drwln{8}{1.0}\spc{2}
                   \drwln{2}{1.0}\spc{2}\drwln{8}{1.0}}\nobreak\ }

\definecolor{darkgreen}{RGB}{0,128,0}
\def\reva#1{ {\color{black}{#1}} }
\def\revb#1{ {\color{black}{#1}} }
\def\revc#1{ {\color{black}{#1}} }
\def\revall#1{ {\color{black}{#1}} }

\graphicspath{{./}}
\def\to{ {\theta_o/U_{o}} }
\def\uo{ {U_{o}} }
\def\lo{ {\theta_o} }
\title[Dynamics of turbulent separation bubbles]
{\color{black}Spatio-temporal dynamics of turbulent separation bubbles} 
\author[Wu, Meneveau and Mittal]{Wen Wu\thanks{Email address for correspondence: w.wu@jhu.edu}, Charles Meneveau, and Rajat Mittal}
\affiliation{
Department of Mechanical Engineering, Johns Hopkins University, Baltimore, MD 21218, USA\\[\affilskip]
}

\pubyear{2010}
\volume{650}
\pagerange{119--126}
\date{?; revised ?; accepted ?. - To be entered by editorial office}

\begin{document}

\maketitle

\begin{abstract}
The spatio-temporal dynamics of separation bubbles induced to form in a  fully-developed turbulent boundary layer (with Reynolds number based on momentum thickness of the boundary layer of 490) over a flat plate are studied via direct numerical simulations. Two different separation bubbles are examined: one induced by a suction-blowing velocity profile on the top boundary and the other, by a suction-only velocity profile. The latter condition allows reattachment to occur without an externally imposed favorable pressure gradient and leads to a separation bubble more representative of those occurring over airfoils and in diffusers. The suction-only separation bubble exhibits a range of clearly distinguishable modes including a high-frequency mode and a low-frequency ``breathing" mode that has been observed in some previous experiments. The high-frequency mode is well characterized by classical frequency scalings for a plane mixing layer and is associated with the formation and shedding of spanwise oriented vortex rollers. The topology associated with the low-frequency motion is revealed by applying dynamic mode decomposition to the data from the simulations and is shown to be dominated by highly elongated structures in the streamwise direction. The possibility of G\"{o}rtler instability induced by the streamwise curvature on the upstream end of the separation bubble as the underlying mechanism for these structures and the associated low frequency is explored.
\end{abstract}


%%%%%%%%%%%%%%%%%%%%%%%%%%%%%%%%%%%%%%%%%%%
\section{Introduction}
Flow separation is ubiquitous in external as well as internal flows: wings and fuselages at high angles-of-attack, flow past external objects, shock-wave/boundary layer interactions, diffusers, corners, inlets and  junctions are just a few examples of this kind of flow. In most of these applications, the incoming boundary layer is turbulent and separates either because of an adverse pressure gradient (APG) or a geometric discontinuity (\textit{e.g.}, backward facing step).  The separated flow typically exhibits unsteadiness across a broad range of time-scales~(\cite{EatonJ82,Cherryetal84, KiyaS85,TenaudPFD16,NajjarB98}, among others). These unsteady modes can dominate the dynamics of the separation bubble and they have significant implications for the performance of the flow device/system at hand. They also introduce difficulties for prediction, 
%of these flow 
but might also offer opportunities for active and passive control of these flows. 

For turbulent separation bubbles, a low-frequency unsteadiness at a time-scale significantly larger than the convective time-scale corresponding to the bubble, is often observed~\citep{TenaudPFD16,NadgeG14, WuMHC05,HudyN97,TaifourW16}. The phenomena has been described as a `breathing' or  `shrinkage and enlargement' of the separation bubble, or a `flapping' of the separated shear layer. \reva{Compared with the high-frequency unsteady mode that represents the vortex generation in the shear layer (and whose time scale is still much larger than the small turbulent eddies' timescales), the low-frequency one is less understood.} Besides its impact on engineering equipment, the low-frequency motion is also a major source of uncertainty in measurements with relatively short averaging times and can cause significant difficulties in the modelling of these flows.
It should be pointed out that most authors have not directly observed a distinctive low-frequency behavior of the separation bubble itself, but they only detected a low frequency in velocity or in wall pressure fluctuations and attributed this frequency to a slight flapping motion of the separated shear layer. In what follows, we will summarize the previous work on the unsteadiness of various separating flows with particular focus on the low-frequency mode. Some of the important questions will be outlined, and the objectives to be pursued in this paper will be presented.

\subsection{Unsteadiness of Separating Flows}
Large-scale unsteadiness is observed in a wide range of configurations that produce flow separation. For geometry-induced flow separation, for example, researchers have examined flow separation at the leading edge of a blunt flat plate~\citep{Cherryetal84}, at the sharp corner of a back-facing step~\citep{{EatonJ82}}  or a diffuser~\citep{KaltenbachFMLM99}, and around a hump~\citep{MarquillieE03} or bluff body~\citep{NajjarB98}. Here, we briefly summarize the previous investigations that have described the unsteadiness quantitatively and proposed some possible mechanisms. 

%\subsubsection{Geometry-Induced Separation}
Among others, \cite{Cherryetal84} and \cite{KiyaS85} conducted experiments on flow separation on a two-dimensional rectangular leading-edge geometry and observed low-frequency motions. \cite{Cherryetal84} described a low-frequency process as a slow modulation of the vorticity shedding from the reattachment zone, wherein the flow goes through  pseudo-periodic shedding of large-scale vortical structures followed by a large-scale but irregular shedding of vorticity, and a relatively quiescent phase with the absence of any large-scale shedding structures. The relaxation time scale between shedding phases is approximately six times of the primary vortex-shedding period. The cause of this low-frequency process was not explained. \cite{KiyaS85} observed a low-frequency process at a similar time scale. Their observations, however, described the low-frequency kinematics as the slight lifting of the shear layer in the vertical direction (less than 4\% of the shear layer thickness) at a low frequency which leads to a change in the vortices that are shed from the separation bubble.  The passage and decay of the varying discrete vortices generate multiple crossings of the zero velocity in the reattachment region and with a short-time averaging it appears as a shrinkage and enlargement of the separation bubble. Following \cite{EatonJ82}, they proposed that the origin of the low-frequency unsteadiness may be the change of the spanwise coherence of the vortices and the variation in their ability in entraining momentum. Some high-fidelity simulations of similar configurations also show the presence of the low-frequency process whose cause was not elucidated ~\citep{TaftiV91,TenaudPFD16}. 

Another widely used configuration for the study of flow separation is the backward-facing step. \cite{EatonJ82} were the first to describe the low-frequency unsteadiness in this flow and they proposed several possible mechanisms for the low-frequency motions. \cite{DurstT82} and \cite{NadgeG14}, among others, support one proposed mechanism, namely of an instantaneous imbalance between shear-layer ``entrainment from the recirculating zone'' and ``reinjection of fluid near reattachment". \cite{HeenanM98}, employed a permeable plate in their experiments and suggested that the low frequency was due to a feeding back of the disturbances from the impingement point to the separation point. Some others attributed the low-frequency motion to the cutting off of recirculation region from the separation bubble by a large structure which reaches the wall well upstream of the mean reattachment point~\citep{McGuineess78,TrouttSN84,DriverSM87,Hasan92}. Other mechanisms that have been proposed include the growth and breakdown of the secondary recirculation region near the lower step corner~\citep{Spazzinietal01,HallBM03}, and the absolute unstable mode of the recirculation region~\citep{WeeYAG04,HudyN97}.  It is clear from this short and by no means exhaustive review that not only is a commonly accepted physical explanation for the low-frequency motion unknown presently, \reva{but evidence about whether a specific observed behaviour is a cause or a consequence of the unsteadiness is still lacking.}

%\subsubsection{Pressure-Induced Separation in Incompressible flows}
\reva{There are also numerous studies on the unsteadiness in the wake of the bluff bodies that have implications for this research. Due to the complex vortex shedding in the wake, the unsteadiness and occurrence of the low-frequency motion is usually explained by the imperfect synchronization and interaction between different types of vortical structures present in the flow~(\cite{GerichE82}, \cite{NajjarB98}, and \cite{WuMHC05} among others). In general, a geometry-induced flow separation is configuration dependent and the mechanisms of the unsteadiness may vary from one configuration to another.} 

\reva{Flow separation can also be induced by an adverse pressure gradient (APG) without the presence of a surface divergence. Compared with the geometry-induced fixed-point separation, a pressure-induced separation is not influenced by the surface curvature, thus being less configuration dependent. Many investigations on flow separation fall in the category of APG induced separation. Before reviewing previous research on pressure-induced separation, we will briefly summarize flow separation over airfoils, which is caused by a combination of a mild surface curvature and an APG. Because airfoils are streamline-shaped, the change in geometry is relatively moderate and the separation point is usually not at a fixed position, e.g. a sharp leading edge or step corner. 
}

% short bubble & long bubble
The study on flow separation over airfoils originates back to the observations of \cite{Jones34} and mainly focuses on laminar incoming flow. One critical feature of laminar flow separation over airfoils is an abrupt increase in the bubble length that occurs occasionally. \cite{OwenK53} classified the closed separation region into `short' and `long' bubbles. The former one were found where the length of the bubble is of the order of 1\% of the chord length and $O(10\delta^*_s)$ to $O(1000\delta^*_s)$, and `long' bubbles with length of the order $O(10,000\delta^*_s)$ ($\delta^*_s$ being the displacement thickness at the point of separation). The `bursting' from short to long bubble crucially impacts the aerodynamic performance of the airfoil because the short bubble only affects the pressure distribution locally while a long bubble can alter the overall pressure distribution around an airfoil~\citep{Tani64}. Parameters such as local Reynolds number or pressure gradient in the region of the bubble~\citep{Gaster63}, as well as semi-empirical models~\citep{Horton68} have been proposed to characterize the bubble bursting. While most studies have found that the bursting into long bubble is caused by a small change in the incidence or speed, more recent investigation indicates the possibilities of an absolute instability that cause the bursting~\citep{AlamS00,JonesSS08}, or an acoustic feedback mechanism~\citep{JonesSS10}, or a coupling of viscous-inviscid interaction and freestream disturbances~\citep{MarxenR10}. It is still an open question as to whether bubble bursting is driven by a change in the stability characteristics of the bubble or by some global instability of the flow. The low-frequency `flapping' motion in the aft part of the separated shear layer is also reported in some studies on laminar separation bubbles (LSBs) over airfoils but the reason is unknown~\citep{HainKR09}.  

To examine flow separation in the absence of surface curvature, experiments have employed the use of aspirated boundaries or contoured ceilings (either concave diverging-converging~\citep{PerryF75,Patrick87,WeissTS15,TaifourW16} or convex converging-diverging~\citep{MarxenRS03,MichelisYK18}) configurations. Simulations have employed a variety of suction-blowing-type boundary conditions
on the top boundary of the computational domain~\citep{NaM98a,Abe17,WuPiomelli18,SpalartStrelets00,AlamS00,MarxenH10,Kotapatietal10,Seoetal18}. 
% LSB
For LSBs, it has been reported that the disturbance amplification agrees well with linear-stability theory, often with an instability of the convective Kelvin-Helmholtz (KH) type. This is followed by a sudden breakdown to three-dimensional, small-scale turbulence~\citep{SpalartStrelets00,RistM02,MarxenH10}. Unsteadiness of the flow is mainly characterized by the shedding of spanwise rollers. \cite{PauleyMR90} were the first to attempt to simulate two-dimensional LSBs ($Re_\theta$ from 162 to 325). They found that some bursting reported in early studies (\cite{Gaster63} among others) is actually periodic shedding which has been time-averaged. \cite{SpalartStrelets00} reported a slight flapping of the separated shear layer in their study of a laminar separation bubble at $Re_\theta=180$ but claimed that the three-dimensionality sets in rapidly, thereby not allowing time for the development of other longer time-scale unsteady motions. 
% TSB

Recently, Weiss, Mohammed-Taifour and co-workers investigated the unsteady behaviour of a massively separated turbulent separation bubble (TSB) generated by a combination of APG and favourable pressure gradient (FPG) ($Re_\theta=5000$). In addition to broadband fluctuations associated with turbulence, the flow displayed a low frequency ``breathing'' or ``flapping" mode and a high frequency ``shedding" mode~\citep{WeissTS15, TaifourW16}. The frequencies of the two modes are centered at Strouhal number $St=fL_{sep}/U_\infty$ of 0.01 and 0.35 ($L_{sep}$ is the length of the mean separation bubble), respectively.   
The shedding mode, as in the geometry-induced separations, can be characterized quite well by the KH instability of the separated shear layer. The origin of the low-frequency mode, however, is less clear. The authors mentioned that the counter-rotating vortices near the sidewalls of the tunnel may affect the unsteadiness. The frequency ranges they reported are close to the one in the literature.
\cite{HudyN97} compared the frequencies observed in different experiments on geometry-induced separation and showed that they are nearly constant: $St\approx$ 0.08-0.2 for the flapping motion and 0.5-1.0 for the shedding motion. It is however unclear as to why the breathing mode appears to have a frequency that is about 10 times lower than the one in fixed-separation flows.

Simulations provide the capability to examine the spatio-temporal dynamics of these turbulent separation bubbles in a way that is difficult to accomplish in experiments. The first DNS of a TSB was performed by \cite{NaM98a} with a $Re_\theta=300$ and they observed that both the separation and reattachment were highly unsteady.
However, these simulations did not exhibit any low-frequency motion and neither have any of the subsequent ones at $Re_\theta=$300, 600 and 900~\citep{Abe17}, \revb{at $Re_\theta = 500-1500$~\citep{ColemanRS18}}, and at $Re_\theta=2500$~\citep{WuPiomelli18}. 

In order to address this gap, we conduct DNS of two types of separation bubble induced on a turbulent boundary layer developing on a flat plate; the first is a separation bubble similar to these previous studies, where separation is induced by applying a suction-blowing velocity on the top boundary of the computational domain. The second bubble is induced by a velocity profile that has suction only. 

\revc{Flow separation is also a topic that has been studied extensively in the high-speed flow community. Two shock-induced separation configurations that have been extensively used in research are the compression ramp and the reflection of an incident shock wave impinging on the boundary layer. Low-frequency unsteadiness in both settings appears as a large-scale motion of the shock. There remain some outstanding questions and debates about the source and mechanism of the low-frequency motion in shock-induced separations~\citep{DussaugeDD06,TouberS09,ClemensN14}. It has been shown that the source of the shock oscillations could be either from the upstream condition or the unsteady recirculating zone.  There are however fundamental differences between the current TSB and  shock-induced separation including the presence of wave propagation associated feedback within the separated flow region and  forced reattachment due to downstream shocks in the latter, and the relative shortness of shock-induced bubbles (around 40$\theta_o$ to 100$\theta_o$ compared to 100-10000$\theta_o$ for pressure gradient induced TSBs).}

\subsection{Motivations and Objectives}
% suction-only configuration
The vast majority of computational studies and many experimental studies of APG induced separation have employed a suction-and-blowing configuration~\citep{PerryF75,Patrick87,NaM98a,WeissTS15,Abe17,WuPiomelli18,WuSMM18}.  An advantage of this configuration is that the total mass flux at the inflow and outflow planes remains the same and this might simplify the the experiments as well as the analysis. However, this type of configuration generates an APG followed by an FPG and the latter leads to a forced reattachment of the flow. However, natural separating flows such as on airfoils and in diffusers do not have this type of forced FPG driven reattachment and we expect that this would also impact any low frequency modes (breathing/flapping) that involve the opening and closing of the bubble. Motivated by these expectations, we initially compare a suction-blowing (or APG-FPG) induced separation bubble with suction only (APG only) induced separation bubble at the same Reynolds number. Subsequently, our attention focuses exclusively on the latter configuration since it is more representative of separation flows in most applications and we conduct a detailed analysis of the unsteadiness and spatio-temporal structure of this `natural' TSB. 

%------
\reva{The suction-only pressure distribution has been used in previous experiments~\citep{SongDE00,DandoisGS07,DianatC91,AlvingF96} and simulations  ~\citep{PauleyMR90,AlamS00,SpalartStrelets00} before, many of which have focused on laminar separation bubbles ~\citep{PauleyMR90,AlamS00,SpalartStrelets00}.  The mean features of the separation bubble have been extensively reported but less attention has been given to the dynamics of suction-only separation bubbles or the low-frequency ``breathing" or ``flapping" behavior. The present study on turbulent separation bubbles focuses on the low-frequency behavior and provides detailed comparisons between the dynamics of suction-blowing and suction-only separation bubbles.}

%------
In the present DNS of natural TSB, a massive, long separation bubble is induced and the reattachment of the mean flow is due to the turbulent diffusion of momentum. Special attention is given to the unsteadiness of the separated flow. In particular, we examine the different time scales of the unsteadiness near the separation and reattachment point both in the vicinity of the wall and in the separated shear layer. We discuss the mechanisms that determine the unsteadiness by relating them to certain types of flow instability. 

In the following, we first present details of the numerical methodology, and then discuss the mean field features, as well as the instantaneous flow evolution. We then describe the unsteadiness of the separation bubble and discuss possible underlying mechanisms. Finally we draw the main conclusions and make recommendations for future work.

\section{Methodology}
\subsection{Configuration}
\begin{figure}
\centering
  \includegraphics[width=0.95\textwidth]{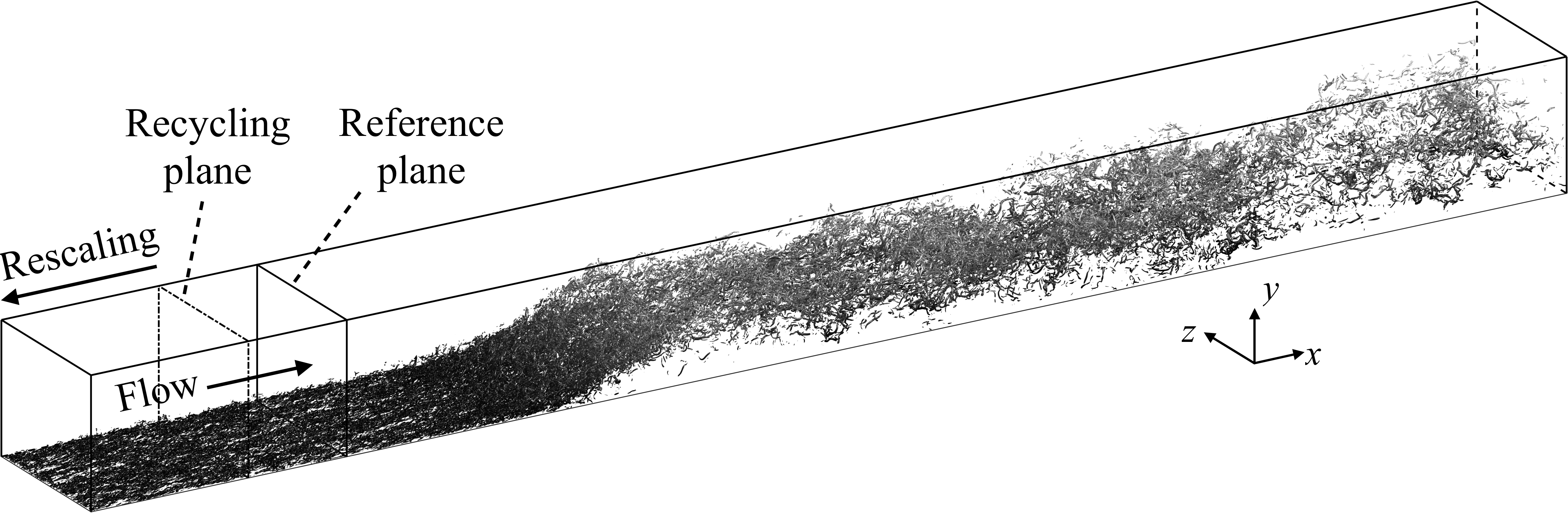}
  \caption{Computational domain. An instantaneous flow field of the TSB-SO case is shown in the domain, visualized by the isosurfaces of the second invariant of the velocity gradient tensor $Q = - \frac{1}{2}\frac{\partial u_j}{\partial x_i}\frac{\partial u_i}{\partial x_j} = 0.0165 U_o^2/\theta_o^2$. Note that $x$, $y$ and $z$, are respectively, the streamwise, spanwise and wall-normal directions.
}
  \label{fig:domain}
\end{figure}

We perform direct numerical simulations of a turbulent boundary layer at $Re_{\theta} =  U_o\theta_o/\nu = 490$, induced to separate via an APG generated by suitable application of a velocity boundary condition on the top of the computational domain.  
%-----------
\reva{This Reynolds number is comparable to recent DNS of separating TBLs (e.g., \cite{ColemanRS18} at $Re_\theta$ = 500 - 1500 in the ZPG region, and \cite{Abe17} at 300, 600 and 900). The Reynolds number is limited by the resolution demands imposed by the very long separation bubble that develops in this flow and the particularly long simulation time required to characterize the low-frequency motion.}
%-----------
The subscript $(.)_o$ refers to scales used for non dimensionalization, namely the freestream streamwise velocity and the momentum thickness at a streamwise location upstream of the APG region. The corresponding $Re_\tau= u_{\tau,o}\delta_o /\nu$, where $\delta_o$ is the boundary layer thickness at the reference plane, is $Re_\tau=200$. The domain size, grid points and resolution used are listed in Table \ref{tab:cases} and the computational domain is shown in Fig. \ref{fig:domain}.  
\begin{table}
  \begin{center}
    \def~{\hphantom{0}}
    \begin{tabular}{lcccccccccc}
      Cases &
      $Re_{\theta,o}$ & $Re_{\tau,o}$ & $Re_{\delta,o}$ &  {$[L_x \times L_y \times L_z]/\theta_o$} & {$N_x \times N_y \times N_z$}  & 
      $\Delta x^+_o$ &  $\Delta y_{o,\text{min}}^+$ & $\Delta z^+_o$ \\[3pt]
      TSB-SB & 490 & 200 & 4150 & $\;935\times100\times117$ & $2304\times408\times384$ &  9.6 & 0.58  & 7.2 \\   
      TSB-SO & 490 & 200 & 4150 & $1170\times100\times117$  & $3072\times408\times384$ &  9.0 & 0.58  & 7.2 \\
      \end{tabular}
   \caption{Simulation parameters. $N_{i}$ is the number of  grid
     points in the entire domain.}
    \label{tab:cases}
  \end{center}
\end{table}

\subsection{Turbulent Boundary Layer Upstream of the TSB}
The recycling and rescaling method by \cite{LundWS98} is utilized to generate the inflow turbulent boundary layer. We also employ a constant spanwise shift of $L_z/2$ for the fluctuating components at each time step to reduce the streamwise correlation of the turbulence between the inlet and the recycling plane~\citep{SpalartST06}. In the current configuration the distance between the inflow and the recycling plane is greater than 17 $\delta_\text{rcy}$, or $125\,\theta_o$, larger than the minimum 11 $\delta_\text{rcy}$ recovery length suggested by~\cite{MorganLKL11} to avoid spurious periodicity. 
\revb{We use a large spanwise domain to prevent an artificial constraint to the turbulent structures in the separated shear layer~\citep{AbeMMS12,AsadaK18}. A doubling of the domain size in the spanwise direction shows negligible changes in the first- and second-order statistics of the flow (Fig. \ref{fig:gridconv}). We have also verified from shorter spatio-temporal maps (not shown) that the key features of the unsteadiness of the separation bubble, including the low frequency, also remain unchanged.}

\subsection{Boundary Conditions}
\begin{figure}
\centering
  \includegraphics[width=0.8\textwidth]{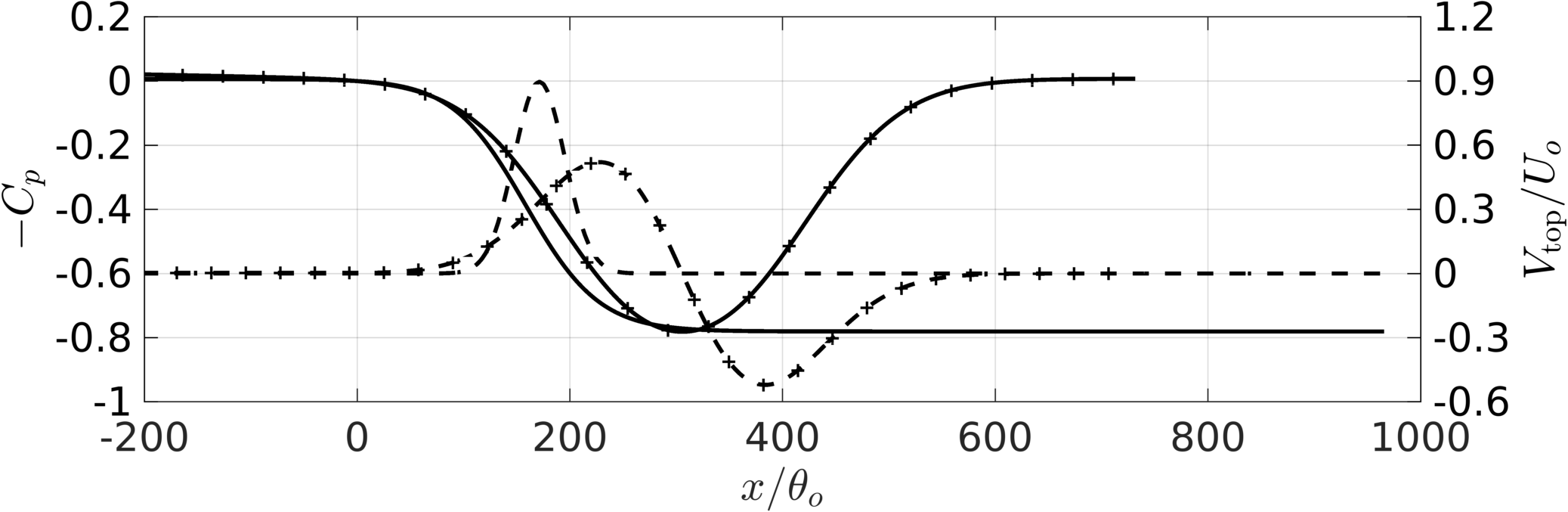}
  \caption{Profiles of the vertical velocity at the top boundary (\dashed) and inviscid $C_p$ at the bottom boundary (\solid). Lines with markers, TSB-SB; lines only, TSB-SO. }
  \label{fig:vtop}
\end{figure}

A convective boundary condition is used at the outflow and periodic conditions are employed at the spanwise boundaries. 
At the top boundary, a non-zero vertical velocity is prescribed so as to generate an APG in the domain while the streamwise velocity satisfies zero vorticity and the spanwise velocity component has zero vertical gradient. Two different vertical velocity profiles are used in this study. The first is a suction-blowing (SB) type of profile which is similar to the profile used in \cite{Abe17} and others, and is given by
\begin{equation}
V_\text{top} = V_\text{max}\sqrt{2}\left(\frac{x_c - x}{\sigma}\right)\exp\left[\psi -\left(  \frac{x_c - x}{\sigma}\right)^2\right]
\end{equation} 
with $V_\text{max}/\uo=0.3325$, $x_c/\lo = 10^6/2^{12} + 62.5$, $\sigma/\lo = 100\sqrt{2} \times 5^2/2^5$ and $\psi=0.95$.
This profile corresponds to a suction followed by blowing  and leads to an APG followed by an FPG on the bottom wall, forming a inverted bell-shaped inviscid pressure profile on the lower wall (see Fig. \ref{fig:vtop}). The net mass flux over the entire top boundary is zero. This case will be referred as `TSB-SB' hereafter and used as a reference for comparison.

The other, suction-only (SO), profile is obtained by aiming to impose a  pressure distribution over the bottom wall that has an APG profile that is similar to that over a typical airfoil section near stall. In particular, we use the pressure distribution of inviscid flow over the suction surface of NACA 0012 airfoil at six degree  angle-of-attack as a guide to obtain a suction velocity profile on the top surface. We employ a two-dimensional, inviscid solver for our rectangular computational domain and employ an iterative approach to adjust the free parameters ($V_\text{max}^\prime$, $x_c^\prime$ and $\alpha$) in the following velocity profile (see \cite{PauleyMR90}, \cite{AlamS00} and \cite{SpalartStrelets00})
\begin{equation}
V_\text{top} = V_\text{max}^\prime\exp\left[-\left(\frac{x-x_c^\prime}{\alpha L_y} \right)^2\right]
\end{equation}
until we reach a APG profile that is a reasonable match to the target. The so obtained values of the profile parameters are $V_\text{max}^\prime/U_o = 0.9$, $x_c^\prime/\theta_o=171.9$ and $\alpha=0.3375$. The resulting profile and inviscid pressure coefficient $C_p=2(P_\text{w} - P_{\text{w},o})/U_o^2$ ($(.)_\text{w}$ denotes quantities at the wall) are also shown in Fig. \ref{fig:vtop}. 
With this suction velocity applied at the top boundary, the nominal deceleration~\citep{PauleyMR90}
\begin{equation}
S = \frac{1}{L_yU_o} \int V_\text{top}(x)dx
\end{equation}
is 0.54. This is much higher than the range that has been presented in previous studies on laminar separation bubbles, \textit{i.e.}, 0.09 to 0.3~\citep{PauleyMR90,SpalartStrelets00}. The value  is in line with the fact that a turbulent boundary layer is harder to separate than a laminar one due to the turbulence-enhanced momentum transfer. Downstream of $x=300\,\theta_o$ we have zero-pressure gradient (ZPG) and allow the separation bubble to develop naturally without any imposed pressure gradient. This suction-only case will be referred as `TSB-SO' hereafter. 
Note that the  start of these two velocity profiles imposed at the top boundary is far downstream of the recycling plane to ensure that the turbulent boundary layer at the recycling plane is not affected by suction. 

\revc{In the current simulation, we employed a very long domain (the domain length is 935 $\theta_0$, although most figure do not show the entire domain) in the streamwise direction and particular attention has been paid to ensure that large-scale structures have decayed significantly near the outflow. It should be noted that in this incompressible flow simulation, there is no mechanism for propagation of spurious reflections from the outflow boundary~\citep{DonG90}. Usually in an incompressible flow simulation, spurious reflected waves accumulate at the exit and lead to numerical instability; this does not happen in the current simulations due to the large domain chosen. Finally,  the flow-through time of a fluid particle between the inflow and outflow is about $1500 \theta_o/U_o$ (note that the mean streamwise velocity is about 0.6$U_o$ after the suction extracts 54\% of the incoming fluid) and this is substantially larger than the time-scale associated with the low-frequency motion, further indicating that the low frequency mode is not associated with the domain size.}  

\subsection{Numerics}
% numerical solver
The calculations are carried out using a well-established flow solver that solves the incompressible Navier-Stokes equations on a Cartesian cell-centered, collocated (non-staggered) grid~\citep{MittalDBNVL08,WuSMM18}. The spatial derivatives are computed using a second-order accurate, central-difference scheme. A second-order Adams-Bashforth scheme is employed for the convective terms and the diffusion terms in the horizontal directions. The diffusion term in the vertical direction is discretized using an implicit Crank-Nicolson scheme that eliminates the viscous stability constraint. The equations are integrated in time using a two-stage fractional step method. The Poisson equation for the pressure is solved by a pseudo-spectral method wherein the three-dimensional Poisson equation is transformed into a set of one-dimensional equations in the wall-normal direction using a Fourier transform in the periodic spanwise direction and a cosine transform in the streamwise direction. Each one-dimensional Helmholtz equation is then solved with the Thomas algorithm for each wavenumber set ($k_x$, $k_z$) before transforming back into the physical space coordinates. The method is second-order accurate in time and space. 

The grid resolutions are listed in Table \ref{tab:cases}. A uniform grid is employed in the streamwise and spanwise direction, and a stretched grid in the wall-normal direction near the bottom wall with a maximum stretching rate of 1.6\%. The grid is also clustered in the wall-normal direction near the upper boundary. For the near-wall flow, the grid size in wall units (wall units are obtained using the local friction velocity $u_{\tau}$) are comparable to the values used in other DNS studies on separating TBL (refer to Table \ref{tab:grid}). Compared with the Kolmogorov scale $\eta$, the present resolution gives $\Delta x/\eta\leq 1.3$, $\Delta z/\eta\leq 1.1$ and $\Delta h/\eta\leq 2$ (where $\Delta h = \sqrt{\Delta x^2 + \Delta y^2  + \Delta z^2}$). Since the maximum dissipation of turbulence occurs at a length scale of about $24\eta$~\citep{Pope2000}, the present grid is able to resolve a substantial part of the dissipation spectrum. 

\begin{figure}
\centering
  \includegraphics[width=0.9\textwidth]{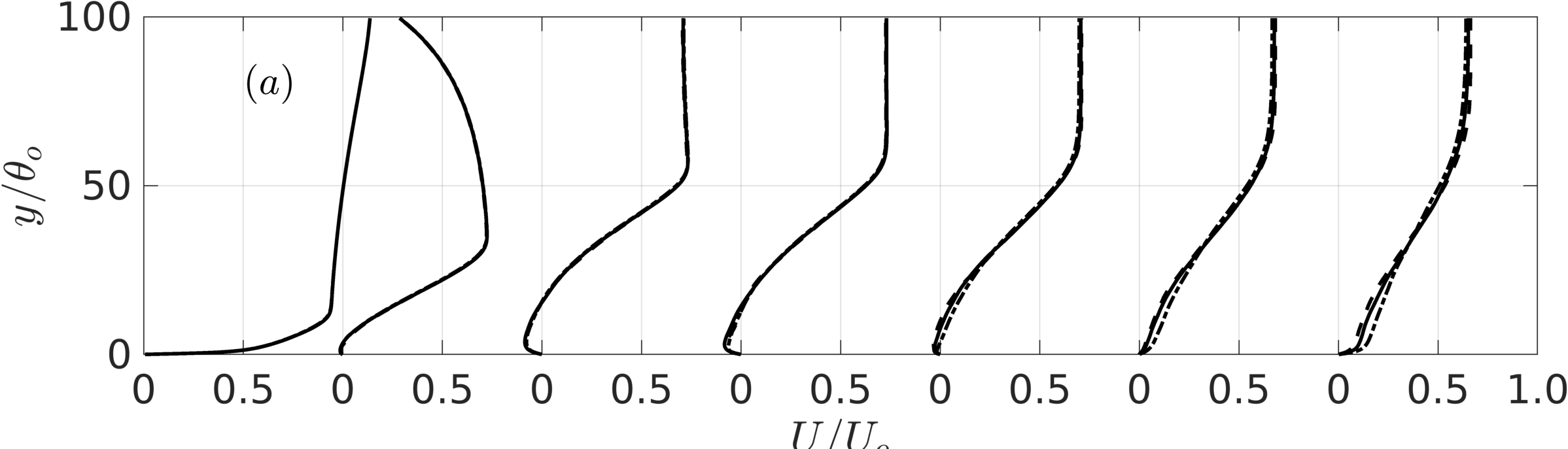}
  \includegraphics[width=0.9\textwidth]{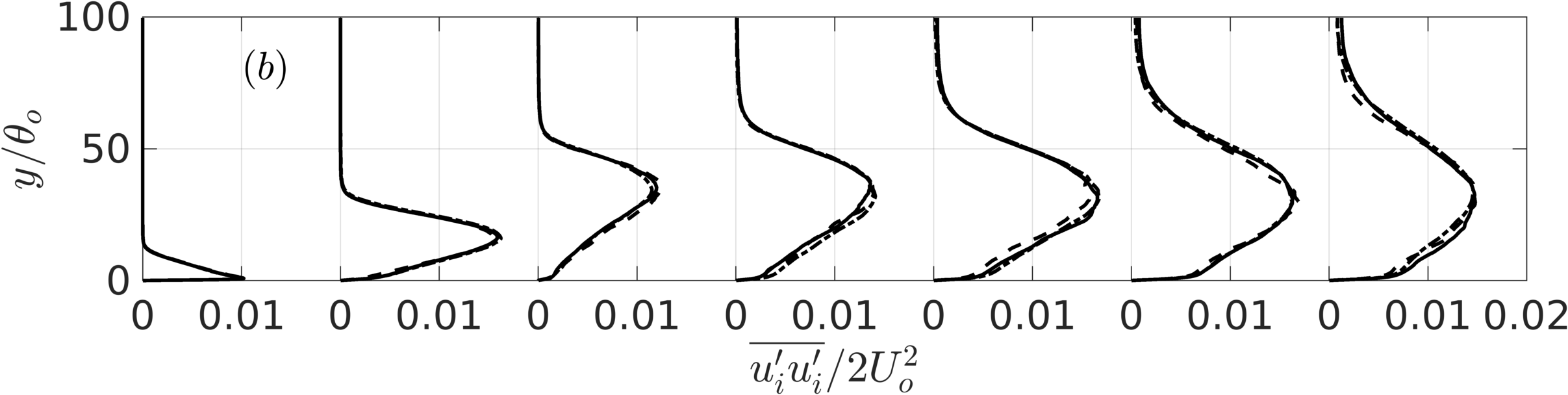}
  \caption{Profiles of mean (a) streamwise velocity, (b) turbulent kinetic energy of the TSB-SO case. Solid line: coarse grid; dashed line: fine grid; \revb{dash-dotted: coarse grid with spanwise domain size doubled.} The examining locations are $x/\theta_o =$ 100, 200, ..., 700 (showing from left to right). Each profile is shifted to the right by one unit in (a) and 0.02 units in (b) for clarity. }
  \label{fig:gridconv}
\end{figure}

\begin{figure}
\centering
  \includegraphics[width=0.47\textwidth]{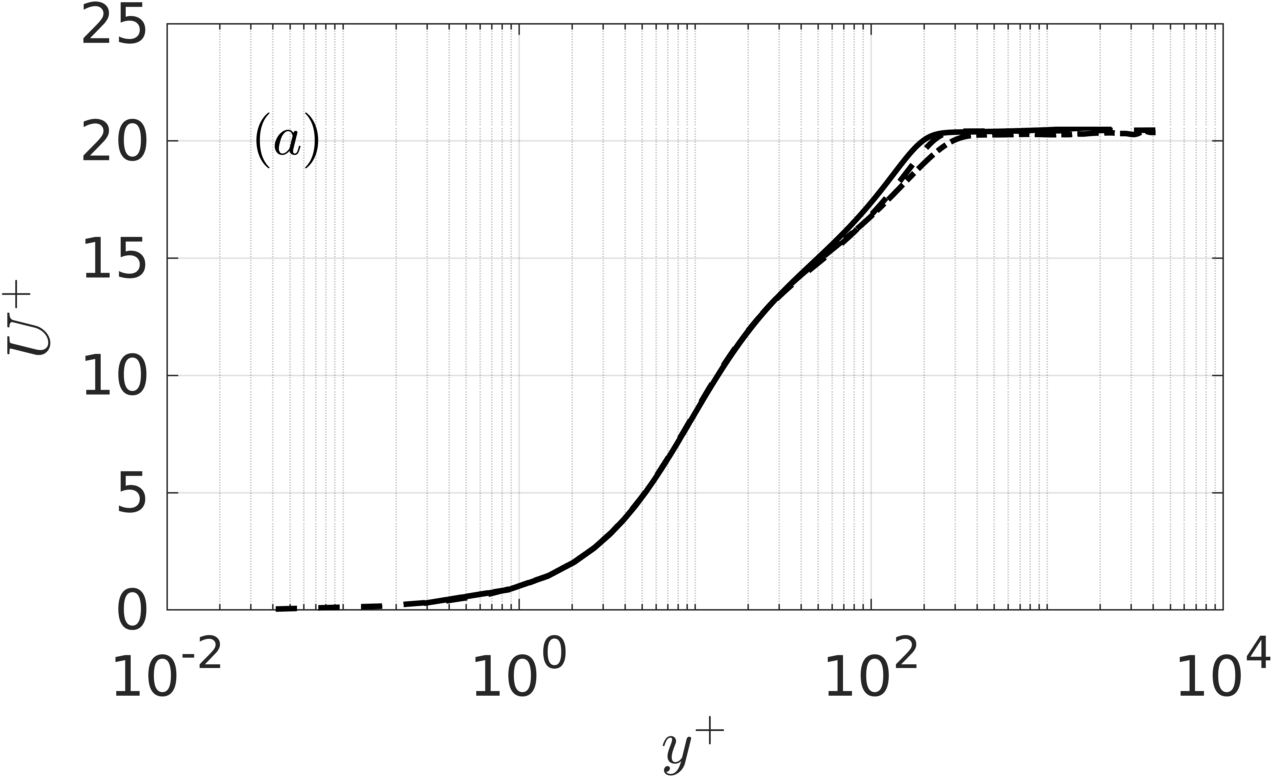}
  \includegraphics[width=0.455\textwidth]{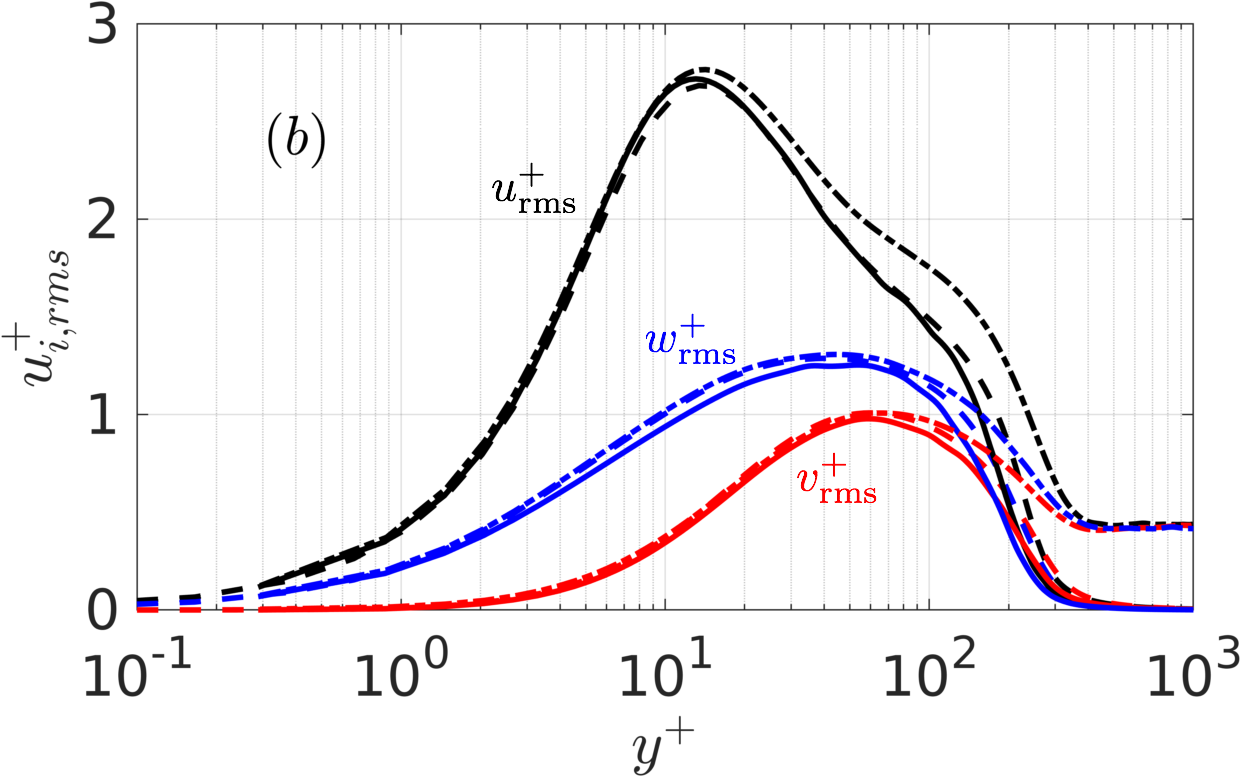}
  \caption{Comparison of (a) mean velocity and (b) Reynolds normal stresses in wall units at $Re_\theta = 670$ for validation. \solid Current results, \dashed \cite{SchlatterO10}, \dashdot \cite{Wuetal17}.}
  \label{fig:valid}
\end{figure}

For the TSB-SO case, a grid that is 20\% finer in each direction (\textit{i.e.}, $\Delta h$ is halved) over the entire calculation domain shows minor difference in the mean velocity and Reynolds stresses, showing that the calculation is grid-converged (Fig. \ref{fig:gridconv}). The code has been validated by performing an
%\textit{a priori} 
initial simulation of a zero-pressure-gradient turbulent boundary layer at $Re_{\theta} = 670$ and comparing the results with previous DNS by \cite{SchlatterO10} and \cite{Wuetal17}. The mean velocity and velocity fluctuations are shown in Fig. \ref{fig:valid} and the agreement is very good.  
\begin{table}
  \begin{center}
    \def~{\hphantom{0}}
    \begin{tabular}{lcccccccccc}
       &$\Delta x/\delta^*_o$ & $\Delta y_\text{max}/\delta^*_o$ & $\Delta z/\delta^*_o$  \\[3pt]
      Present                & 0.24 & 0.25 & 0.19  \\
      \cite{SpalartC97} & 0.57 & 0.27 & 0.21 \\
      \cite{NaM98a}     & 0.85 & 0.41 & 0.48  \\  
      \cite{SkoteH02}   & 0.65 & 0.30 & 0.21  \\
       \cite{Abe17}       & 0.32  & 0.26 & 0.13 \\   \end{tabular}
   \caption{Comparison of grid resolution between present simulation (case TSB-SO) and the previous DNS of TSB.}
    \label{tab:grid}
  \end{center}
\end{table}

% sampling
The data is sampled at a regular time interval of $\Delta t = 3.89\,\to$, once a statistically stationary state was reached. To characterize the unsteadiness at low frequency, the simulation is integrated over a very long duration of $T = 20,000 \,\to$ and statistics are obtained by averaging over time and in the spanwise direction. The difference between the first- and second-order statistics obtained using only half of the sample and the ones using the entire data series is less than 1\%. In the following discussion, the averaged quantities will be denoted via capital letters for primary flow variables, and by $\overline{(.)}$ for turbulent statistics. Quantities that are only averaged in the spanwise direction are denoted by $\langle . \rangle_z$, and other arbitrary averaging by $\langle . \rangle$.

\section{Results and Discussion}
\subsection{Suction-Blowing (SB) versus Suction-only (SO) Separation Bubble}
In this section, we compare the key features of these two separation bubbles. This is followed by a detailed analysis of the unsteadiness in the TSB-SO bubble.
\subsubsection{Vortex Topology}
\begin{figure}
\centering
  \includegraphics[width=0.95\textwidth]{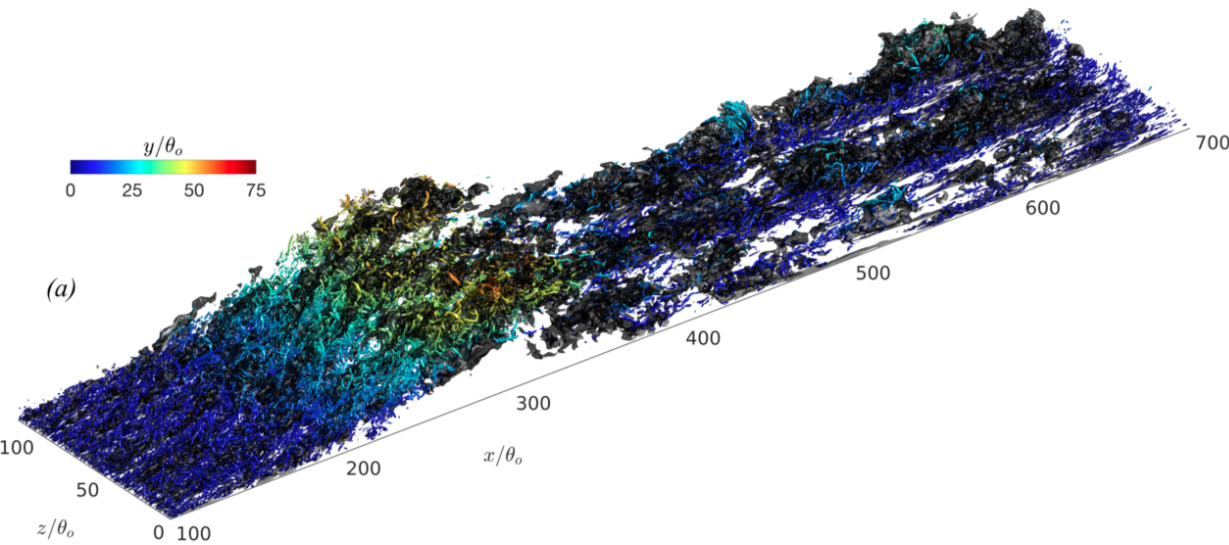}
  \includegraphics[width=0.95\textwidth]{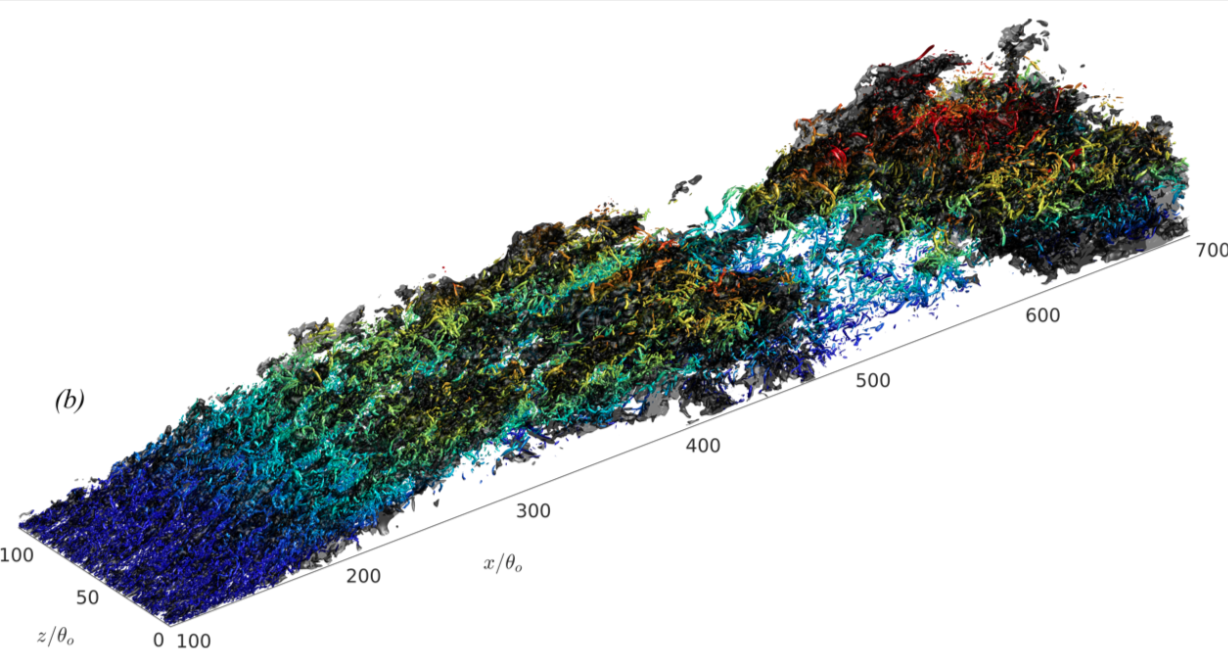}
  \caption{Vortex structures at one time instant for the two separation bubbles. (a), TSB-SB; (b), TSB-SO. Vortex structures are visualized by the isosurfaces of the second invariant of the velocity-gradient tensor $Q=0.0165U_o^2/\theta_o^2$ (see text), colored by the distance from the wall. The dark-gray isosurfaces are $u^\prime=-0.1 \,U_o$. }
  \label{fig:Qiso}
\end{figure}

 Figure \ref{fig:Qiso} shows instantaneous snapshots of the vortex structures in the two separation bubbles via the second invariant of the velocity-gradient tensor
\begin{equation}
Q = - \frac{1}{2}\frac{\partial u_j}{\partial x_i}\frac{\partial u_i}{\partial x_j} = \frac{1}{2}\left(\Omega_{ij}\Omega_{ij} -S_{ij}S_{ij}\right)
\end{equation}
where $\Omega_{ij}$ and $S_{ij}$ are the antisymmetric and symmetric parts of the velocity gradient tensor respectively. It can be seen that many elongated low-speed regions are present in the separated shear layer. The size of these structures in the spanwise direction is relatively small near the separation point and they merge into larger structures as the flow separated. Each of them is surrounded by a group of streamwise-aligned hairpin-like structures. In the TSB-SB case these structures are broken by the blowing, and near-wall streaks form during its recovery to turbulent boundary layer after reattachment. In the TSB-SO cases the structures in the separated shear layer sustain for a long time and break around $x=450\,\theta_o$ at the time-instant shown. A large vorticity packet is observed downstream including small-scale turbulent eddies nested inside the packet.

\subsubsection{Characteristics of the Mean Flow }
\begin{figure}
\centering
  \includegraphics[width=0.9\textwidth]{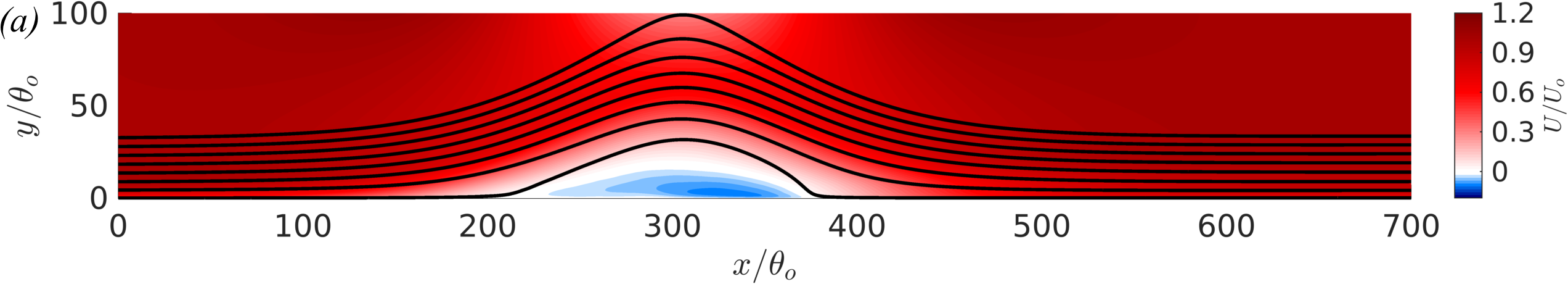}
  \includegraphics[width=0.9\textwidth]{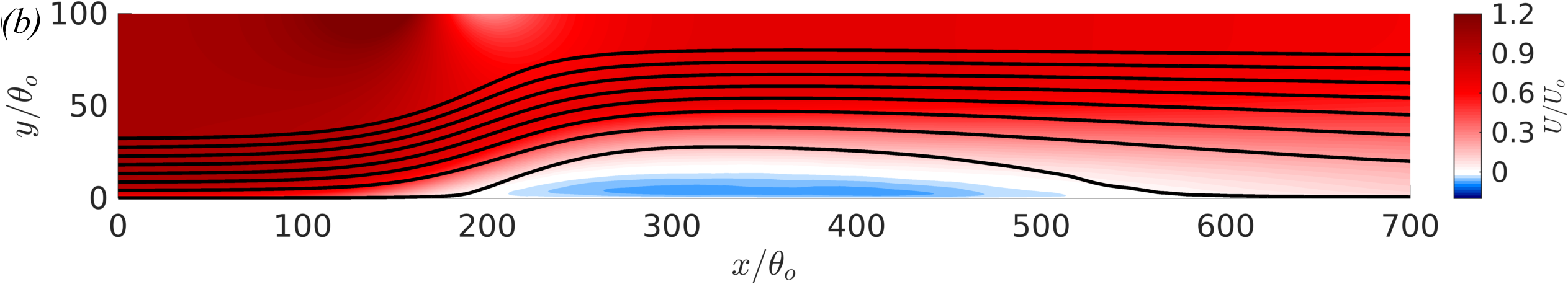}
  \caption{Contours of mean streamwise velocity with selected streamlines (solid). (a), TSB-SB; (b), TSB-SO.}
  \label{fig:umean}
\end{figure}

Figure \ref{fig:umean} shows contours of mean streamwise velocity $U/U_o$ in the $x-y$ plane together with several selected mean streamlines for both the separation bubbles. The characteristic properties of the mean separation bubble are also compared in Table \ref{tab:meanbubble}. The mean separation bubble in the TSB-SO configuration is significantly longer than the one in the TSB-SB case but slightly thinner. Furthermore, the reversed flow in the TSB-SO configuration has a lower magnitude and the streamwise velocity gradient near the reattachment point is also lower due to the absence of the forced impingement of the flow.  The peak reversed flow amplitude, scaled with the far-field velocity at each streamwise location, is about 8\% in TSB-SO and 11\% in TSB-SB. 
\begin{table}
  \begin{center}
    \def~{\hphantom{0}}
    \begin{tabular}{ccccc}
      Cases &
      $x_\text{sep}/\theta_o$ & $x_\text{ratt}/\theta_o$ & $L_\text{sep}/\theta_o$ & $h_\text{sep}/\theta_o$ \\[3pt]
      TSB-SB & 188 & 379 & 191 & 18.5 \\
      TSB-SO & 164 & 614 & 450 & 16.4 \\
   \end{tabular}
   \caption{Characteristics of the mean separation bubble. The mean separation point $x_\text{sep}$ is where $C_f = 2\tau_w/U_o^2=0$ and $d C_f/dx <0$, and the mean reattachment point  $x_\text{ratt}$ is where $C_f =0$ and $d C_f/dx >0$. $h_\text{sep}$ is the maximum distance between the contourline of $U=0$ and the wall.}
    \label{tab:meanbubble}
  \end{center}
\end{table}

\begin{figure}
\centering
  \includegraphics[width=0.9\textwidth]{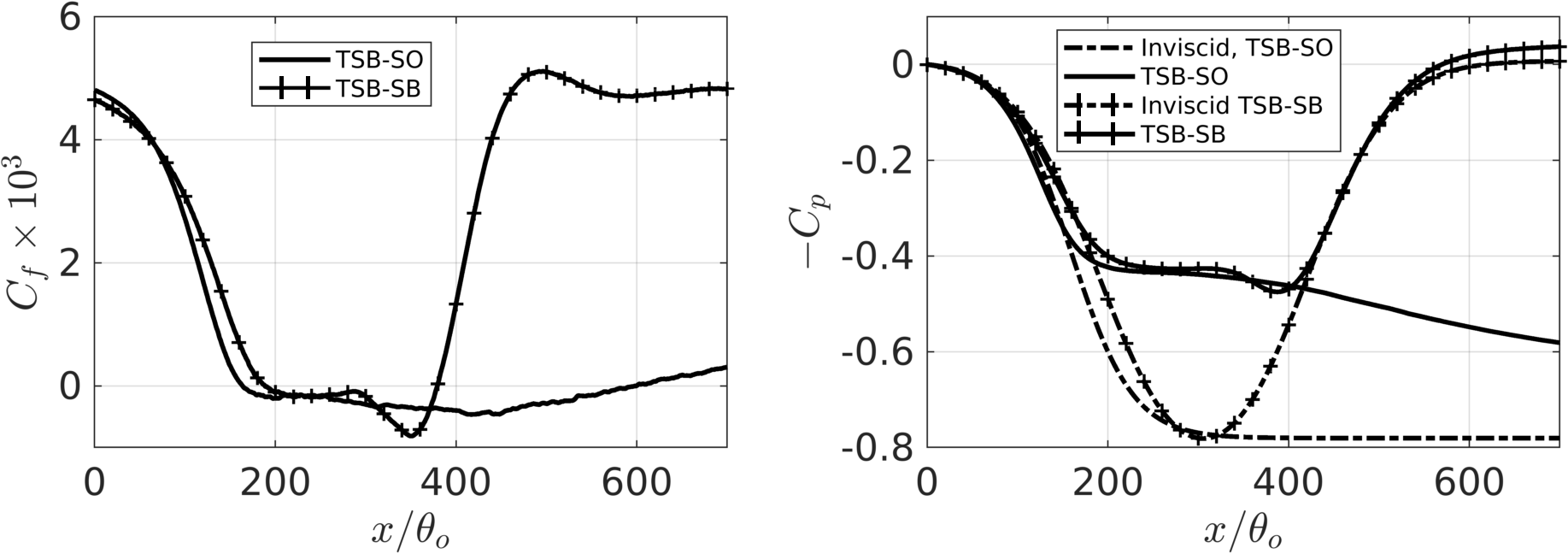}
  \caption{Streamwise profiles of $C_f$ and $C_p$. Lines with markers, TSB-SB; lines only, TSB-SO. Dash-dotted lines in the right subfigure are inviscid wall $C_p$. }
  \label{fig:CfCp}
\end{figure}

The time-mean profiles of skin-friction coefficient ($C_f$) and pressure coefficient on the wall ($C_p$) for both the bubbles are compared in Fig. \ref{fig:CfCp}. In both cases the wall pressure agrees well with that for an inviscid solution up to the separation point. The suction-and-blowing configuration produces a bell-shape inviscid wall pressure profile over the surface that consists of an APG followed by an FPG~\citep{NaM98a,Abe17,WuPiomelli18}. Pressure profiles of realistic separated flow are however, quite different. For an airfoil at large angle-of-attack, a steep APG appears near the leading of the airfoil and gradually decreases towards the trailing edge. However, there is usually no region of FPG ~\citep{RinoieT04,KimYCC09,AsadaK18}. In diffusers with fixed opening angle in the streamwise direction~\citep{KaltenbachFMLM99}, strong APG occurs at the beginning of the  deflected wall and decreases rapidly to ZPG downstream. Unless the wall itself is not parallel to the freestream flow (\textit{e.g.}, a shock-induced separated flow at a compression corner), there is no common mechanism to have a driving mean flow that impinges towards the surface after the flow separates. 

The $C_p$ in the TSB-SO configuration consists only of the APG as expected, and there is no local pressure peak due to the stagnation of the mean flow such as the one observed at $x/\theta_o \sim 390$ in the TSB-SB case. In both cases, the Clauser pressure-gradient parameter
\begin{equation}
  \label{eq:4}
  \beta = \frac{\delta^*}{\tau_w}\frac{dP_{e}}{dx}
\end{equation}
where $(.)_e$ denotes the quantity measured at the edge of the boundary layer, is $\beta = 0.5$ at $x=60\,\theta_o$, $\beta = 3$ at $x=100\,\theta_o$, and then rapidly increases to $\beta = 85$ before $u_\tau$ becomes smaller than 0.01 $U_e$ upstream of the separation bubble. 

\begin{figure}
\centering
  \includegraphics[width=0.7\textwidth]{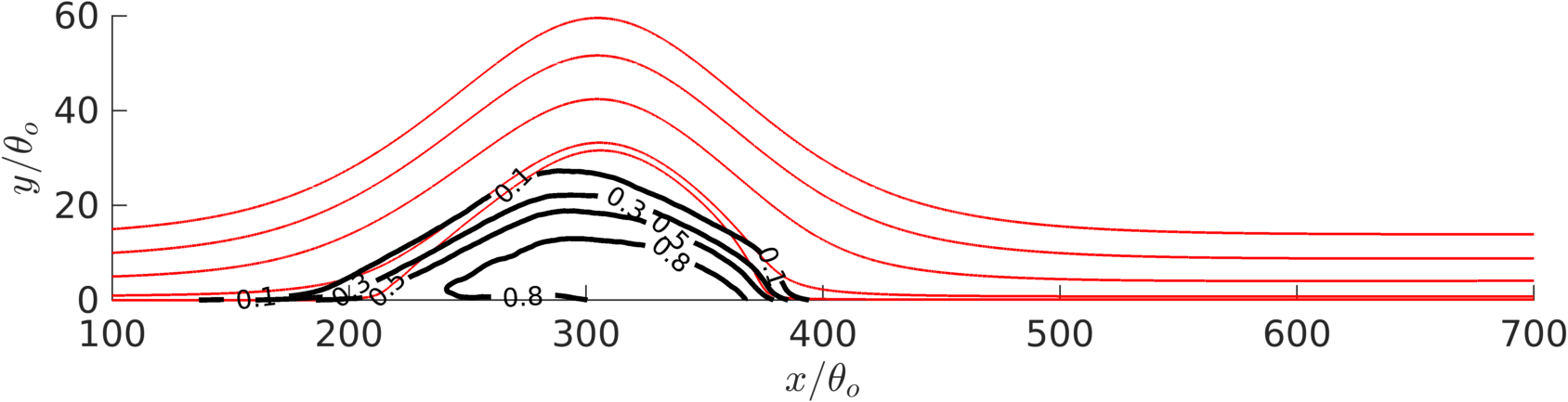}
  \includegraphics[width=0.7\textwidth]{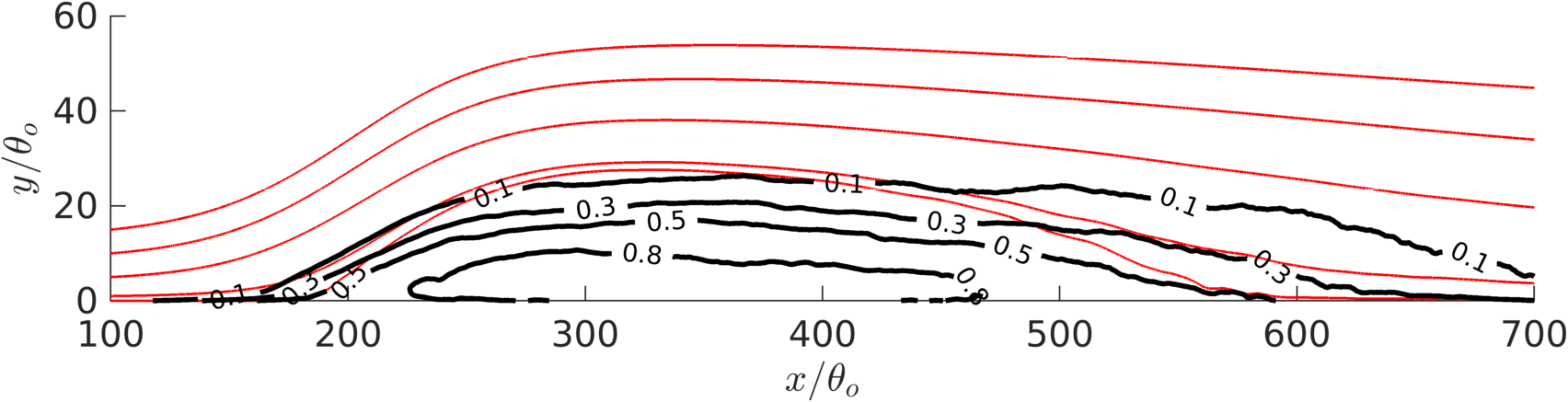}
  \caption{Probability of occurrence of reversed flow as function of position. Top, TSB-SB; bottom, TSB-SO. Light solid lines are selected mean streamlines.}
  \label{fig:unegPDF}
\end{figure}
\begin{figure}
  \centering
  \includegraphics[width=0.45\textwidth]{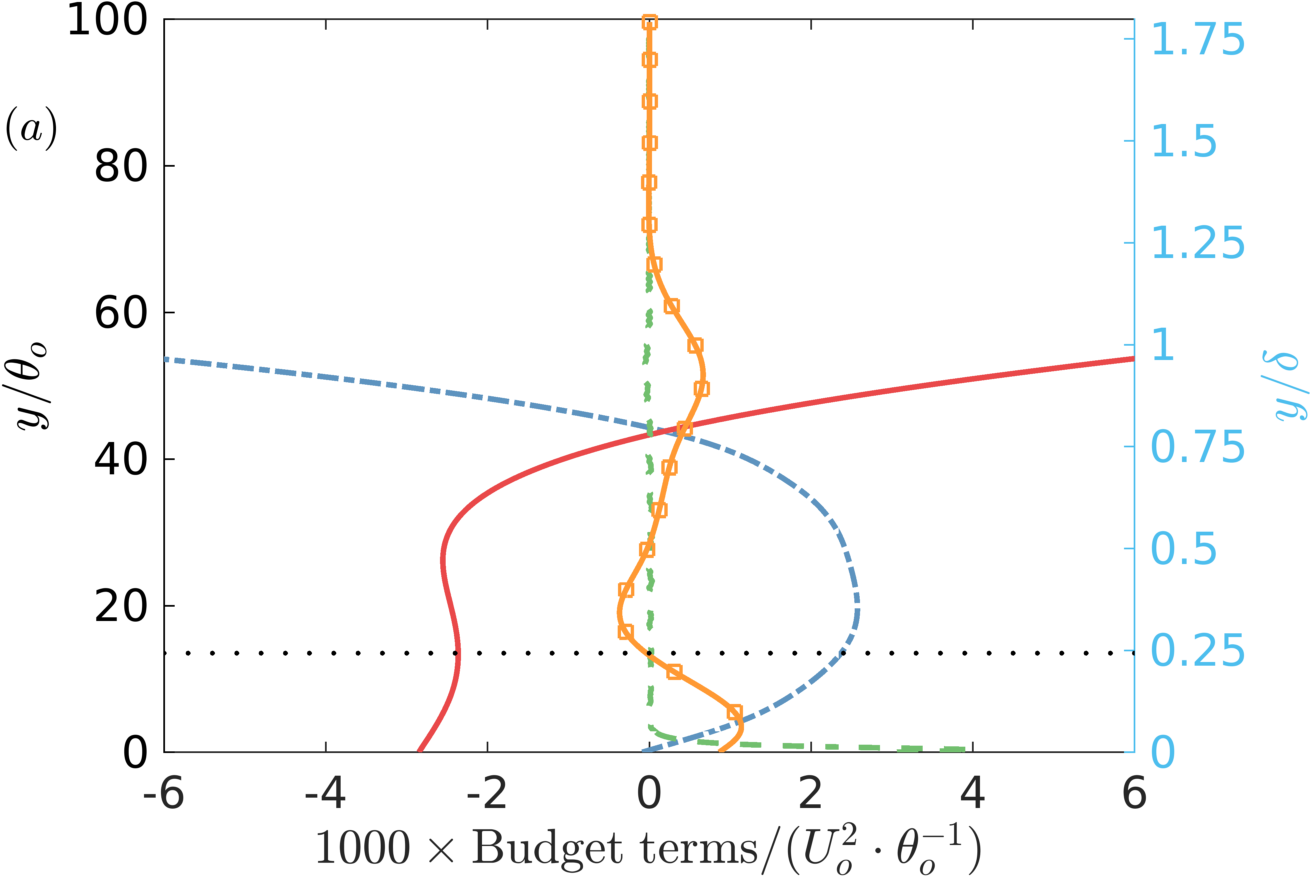}
  \includegraphics[width=0.45\textwidth]{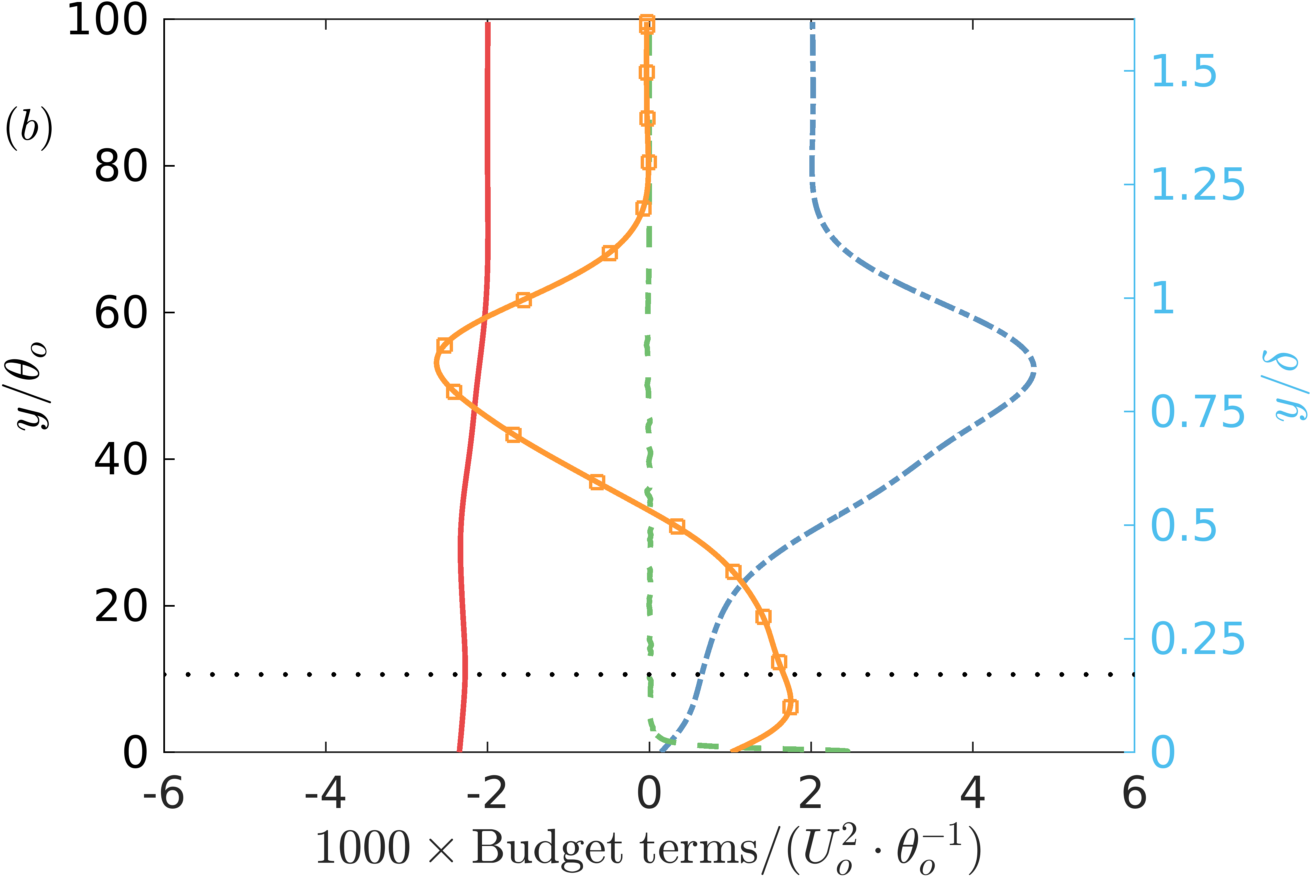}
 \caption{Budget of the mean $x$-momentum balance. Dash-dotted line: advective term $-U_j \partial U / \partial x_j$; solid line, pressure term $-\partial P/\partial x$ ; dashed, viscous term $\nu \partial^2 \overline{U} / \partial x_j^2$; solid with marker, Reynolds stress term $-\partial \overline{u^\prime u_j^\prime} / \partial x_j$. Horizontal dotted line, local $y\vert_{U=0}$. (a) TSB-SB at $x/\theta_o=342$; (b) TSB-SO at $x/\theta_o=467$. The examining $x$-location is chosen as the middle point between the bubble crest and mean reattachment point. The residual in both cases are smaller than $2\times10^{-4} U_o^2/\theta_o$.}
 \label{fig:xmom}
\end{figure}

The probability distribution of reverse flow occurring in the $x-y$ plane is shown in Fig. \ref{fig:unegPDF} for both bubbles. The point of the separation region in both cases shows a very steep gradient in probability, indicating a stationary separation point. However, the reattachment behavior of the two bubble is quite different. In particular, unlike the TSB-SB bubble, for the TSB-SO bubble there is a large region beyond the reattachment point where reverse flow can occur up to 10\% of the time. This is because the reattachment of the mean flow in the case of the TSB-SO bubble is due to the transport of momentum by turbulence and not forced due to the imposed blowing at the top. 
%----------
\revb{In order to confirm this statement, we examine the balance of momentum-flux~\citep{Abe19} in the aft portion (i.e., the region between the bubble crest and the reattachment point) of both separation bubbles (Fig. \ref{fig:xmom}): in the TSB-SB case, the convection term dominates the positive gain in the momentum balance near the $U=0$ `interface', with a small turbulent transport term. In the TSB-SO case, on the contrary, the turbulent transport becomes the leading term in TSB-SO at such interface, and the mean flow only contributes near the edge of the boundary layer.}
%----------
On the wall, in particular, the region between $x/\theta_o = 600$ and $x/\theta_o = 706$ experiences backflow during 50\% to 30\% of the time for the TSB-SO bubble. The extent of this region in the TSB-SB case, in comparison, is only about $6\,\theta_o$ ($x/\theta_o \in [379,\, 385]$). The periodic formation of discrete vorticity packet causes remarkable intermittency in flow reattachment: for 30\% (10\%) of time the bubble is 20\% (40\%) longer than the mean. As will be described later in the paper, this feature of the reattachment is associated with a low-frequency ``breathing" or ``flapping" of the TSB-SO bubble.

\begin{figure}
\centering
  \includegraphics[height=0.26\textwidth]{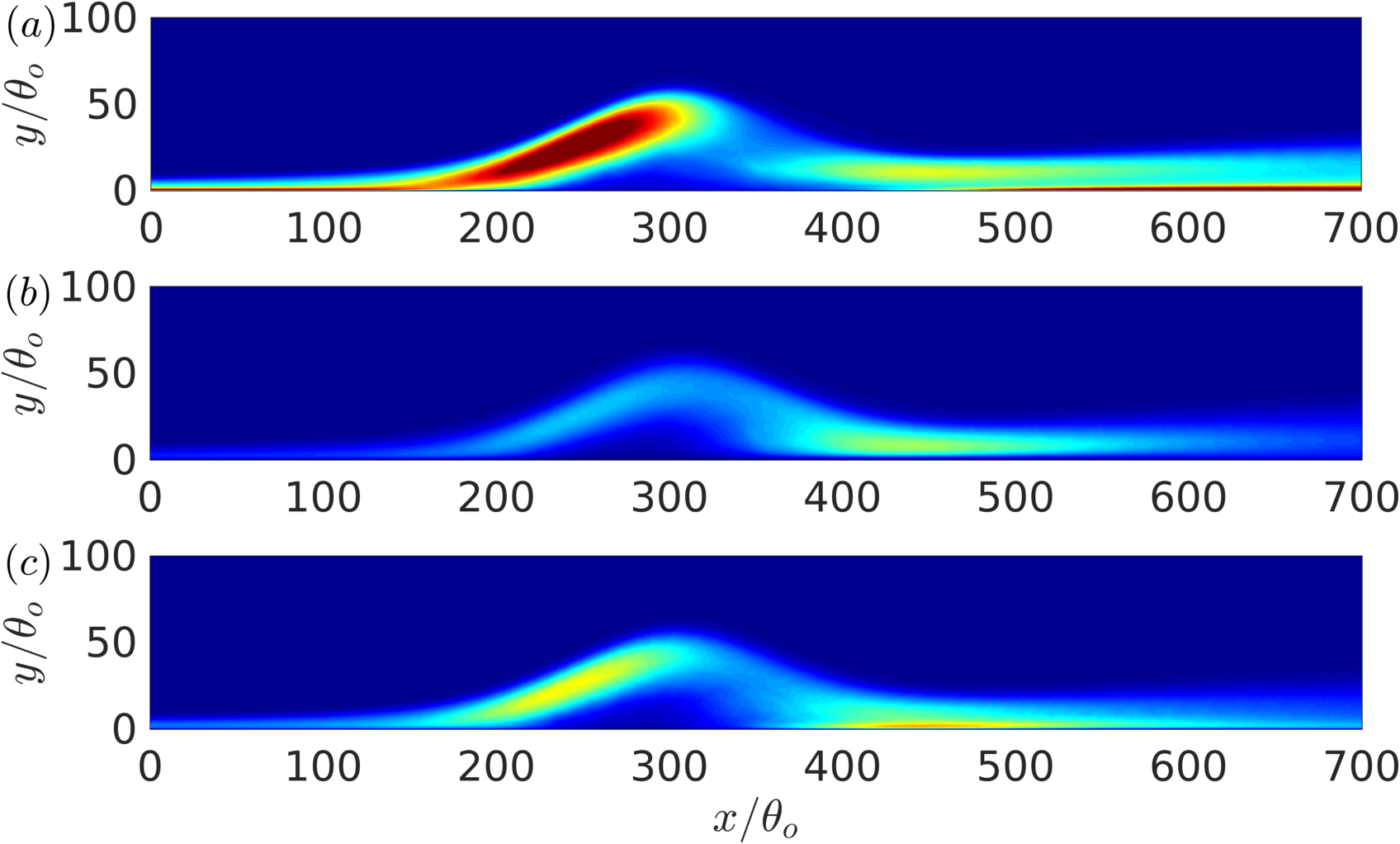}
  \includegraphics[height=0.26\textwidth]{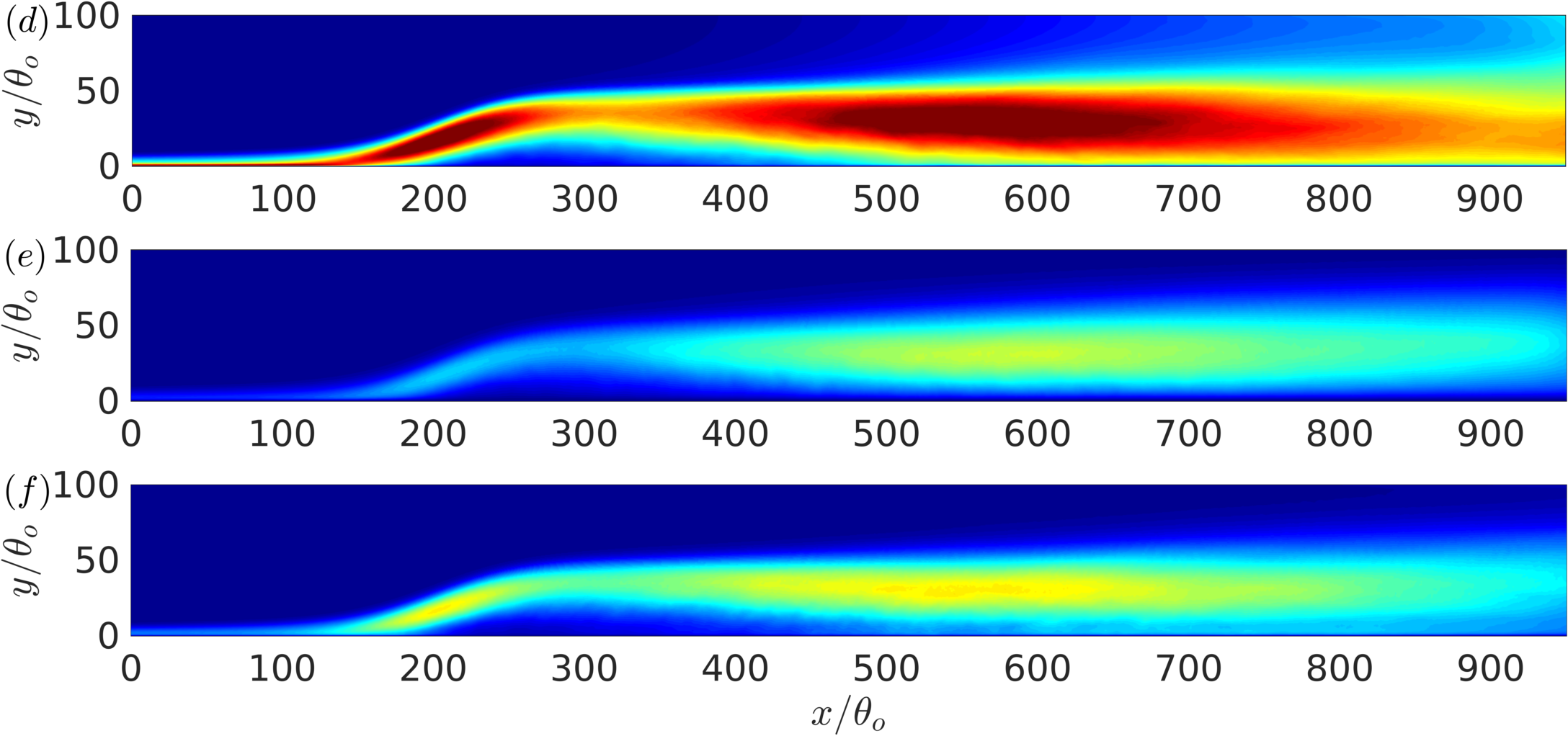}\\
  \includegraphics[height=0.06\textwidth]{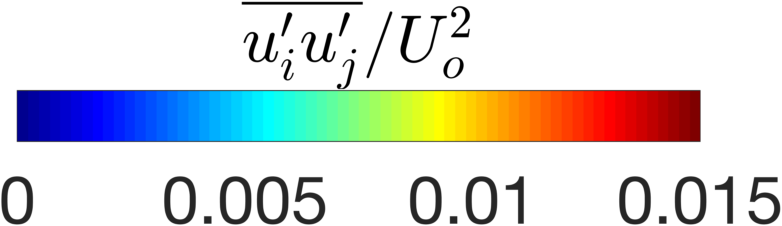}
  \caption{Time- and spanwise-averaged Reynolds normal stresses. (a-c) TSB-SB; (d-f) TSB-SO. Top row, $\overline{u^\prime u^\prime}/U_o^2$; middle row,  $\overline{v^\prime v^\prime}/U_o^2$; bottom row, $\overline{w^\prime w^\prime}/U_o^2$.}
  \label{fig:urms}
\end{figure}

Eliminating the forced impingement due to the imposed blowing also significantly impacts the development of turbulence in the two flows as shown in Fig. \ref{fig:urms}. While the two cases share a similar turbulence profile near the region of initial separation, the TSB-SO case also exhibits another large area of high turbulence down near and above the mean reattachment point. As will be shown later in the paper, this large region of turbulent fluctuations is associated with unsteady motions that occur at a large time and spatial scale. In general, the observed Reynolds stress distributions obtained for the TSB-SO case are in much better qualitative agreement with the ones in flow separation over airfoil~\citep{Balakumar18,JonesSS08} than those for the TSB-SB case. More details about the statistics of the TSB-SB bubble could be found in \cite{WuMM19}.

\subsubsection{Low-Frequency Unsteadiness in the Separation Bubbles}
\begin{figure}
\centering
  \includegraphics[width=0.48\textwidth]{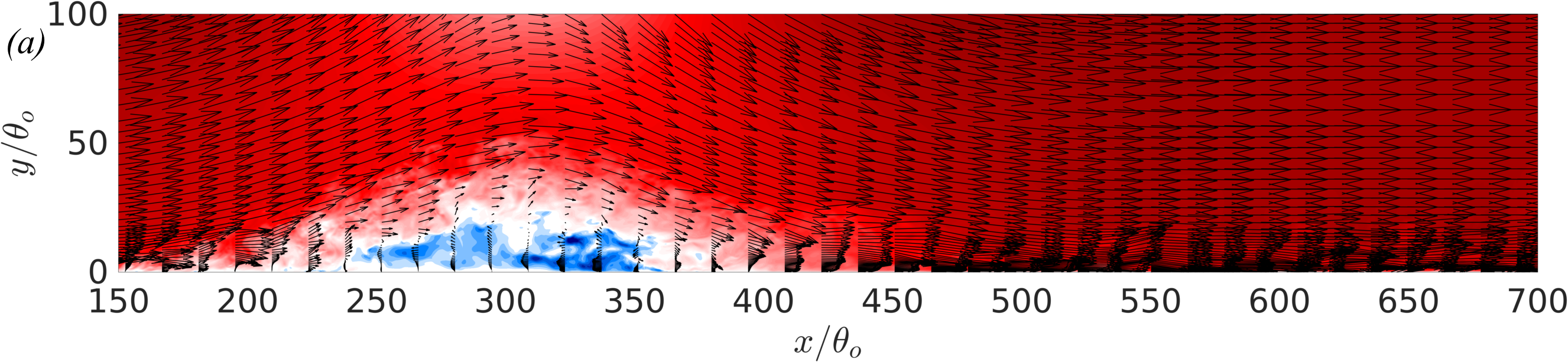}
  \includegraphics[width=0.48\textwidth]{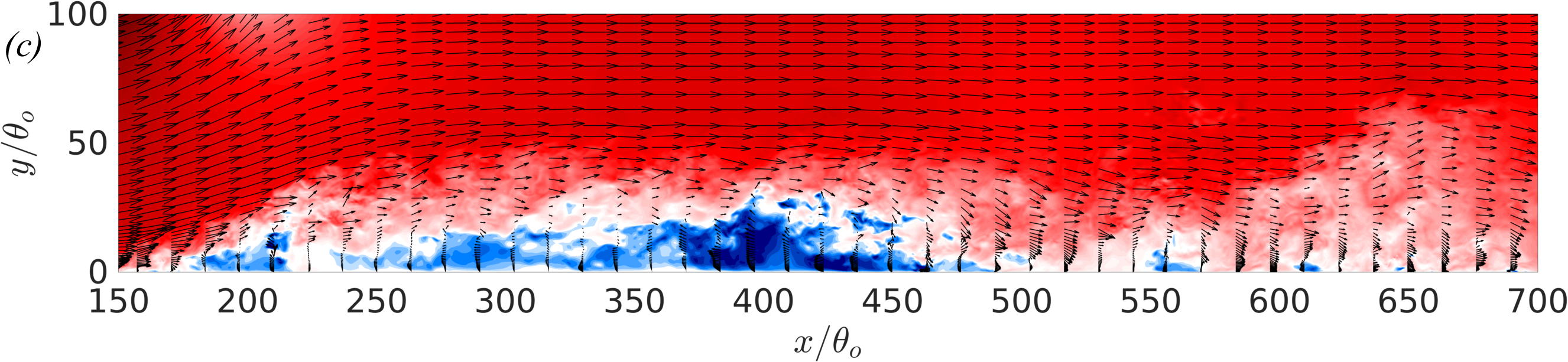}
  \includegraphics[width=0.48\textwidth]{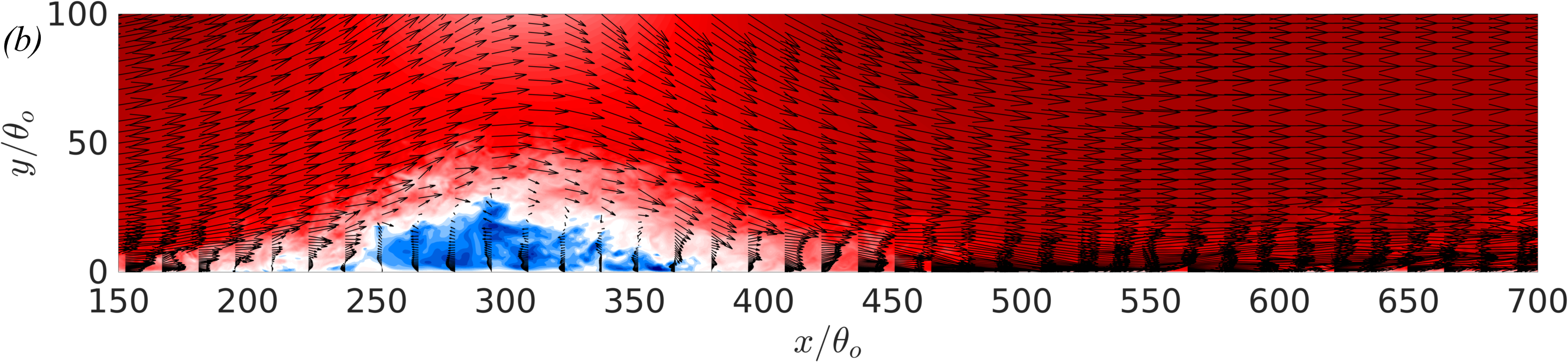} 
  \includegraphics[width=0.48\textwidth]{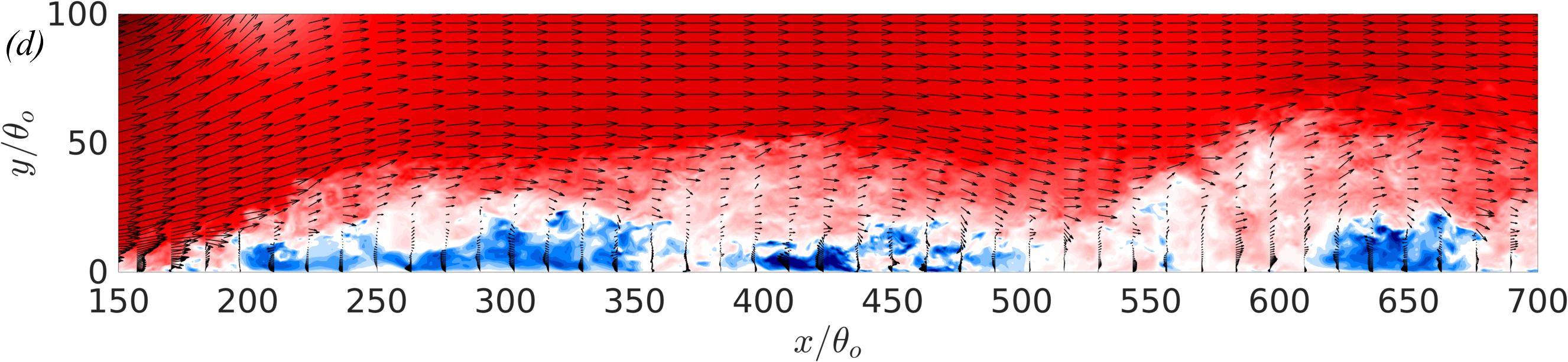} 
  \caption{Instantaneous streamwise velocity and velocity vector in the $x-y$ plane at $z=L_z/2$. (a) TSB-SB, $tU_o/\theta_o \approx$ 9,260, (b) TSB-SB, $tU_o/\theta_o \approx$ 11,170, (c) TSB-SO, $tU_o/\theta_o \approx$ 15,000, (d) TSB-SO, $tU_o/\theta_o \approx$ 16,950. Colormap refer to Fig. \ref{fig:umean}.}
  \label{fig:instU}
\end{figure}

In this section, we examine the two separation bubbles for the presence of a low-frequency  ``breathing" or ``flapping" mode. 
In Fig. \ref{fig:instU} we show contours of instantaneous streamwise velocity and the velocity vectors in the $x-y$ plane at center-span at two different time instants. The TSB-SO bubble shows large variation of shape in time: the bubble is long enclosing a  continuous backflow region (Fig. \ref{fig:instU} (c)) at some times, and has several distinct backflow regions at other times (Fig. \ref{fig:instU} (d)). The TSB-SB bubble on the other hand does not show any such large-scale changes in the size of the bubble with time (Fig. \ref{fig:instU} (a, b)).  This behavior also explains the differences in the probability distributions observed in Fig. \ref{fig:unegPDF} in the rear part of the TSB-SO bubble.

\begin{figure}
\centering
  \includegraphics[width=1.0\textwidth]{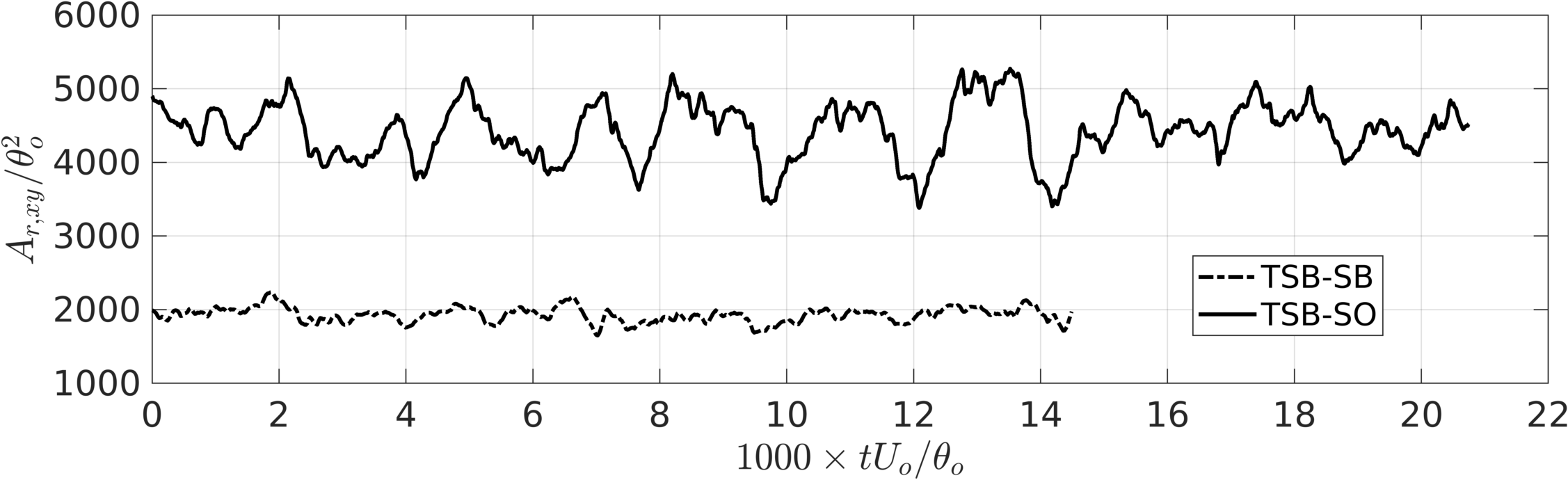}
  \caption{Time history of the total reversed flow area in the $x-y$ plane.}
  \label{fig:Axy}
\end{figure}
The change in the size of the separation bubble versus time serves as a simple parameter that  measures large-scale unsteadiness of a separation bubble. The history of the total reversed flow area in the $x-y$ plane, {\it i.e.}, $A_{r,xy} = \mathcal{\int}_{\Omega_r} dxdy$ where $\Omega_r$ is the region where $\langle u \rangle_z<0$,  is plotted in Fig. \ref{fig:Axy}. A large-amplitude, low-frequency variation of the total backflow area is clearly observed for the TSB-SO bubble whereas a similar behavior is missing the TSB-SB bubble. Thus, taken together, Figs. \ref{fig:instU} and \ref{fig:Axy} clearly suggest that while the TSB-SO bubble exhibits a large-scale low-frequency unsteadiness of the type that has been characterized as the ``breathing" or ``flapping" mode in previous studies, the TSB-SB bubble does not show any such distinct behavior. 
%Given this, and the emphasis of the current study on the low-frequency unsteady behavior of turbulent separation bubbles, we focus the rest of the paper on the analysis of the TSB-SO bubble.

Note that, the low-frequency motion we observed here does not have a very significant separation in time scale with respect to the high frequency ones. As we will describe in detail in the next section, the frequencies of the two motions differ only by a factor of 2-2.5. As summarized in the Introduction, researchers have reported low-frequency motions that occur at a time scale ranging from 6 to 35 times longer than the vortex shedding mode. One reason for this discrepancy may be that we are observing a different type of low-frequency motion in flow separation. It also is possibly related to the relatively low Reynolds number of the current flow since it is expected that while the vortex shedding frequency scales with the inverse of the boundary layer thickness (which decreases with increasing Reynolds numbers), the low frequency mode scales with the size of the separation bubble, which would be less dependent on the Reynolds number. Indeed, in the experiments of TSBs conducted by \cite{WeissTS15} and \cite{TaifourW16}, the boundary Reynolds number was $Re_\theta=5000$ and the observed a low-frequency peak in the spectrum \reva{which was 35 times lower than the vortex shedding peak. Note that, in their experiments, which employed a ceiling with an expansion and mass removal followed by a contraction, there is a FPG following the APG but the net FPG is smaller than the net APG. Thus their configuration falls in between what we define as a SB and SO configurations.}

\subsection{\reva{Characteristics of Unsteadiness in the TSBs}}
\begin{figure}
\centering
  \includegraphics[height=0.55\textwidth]{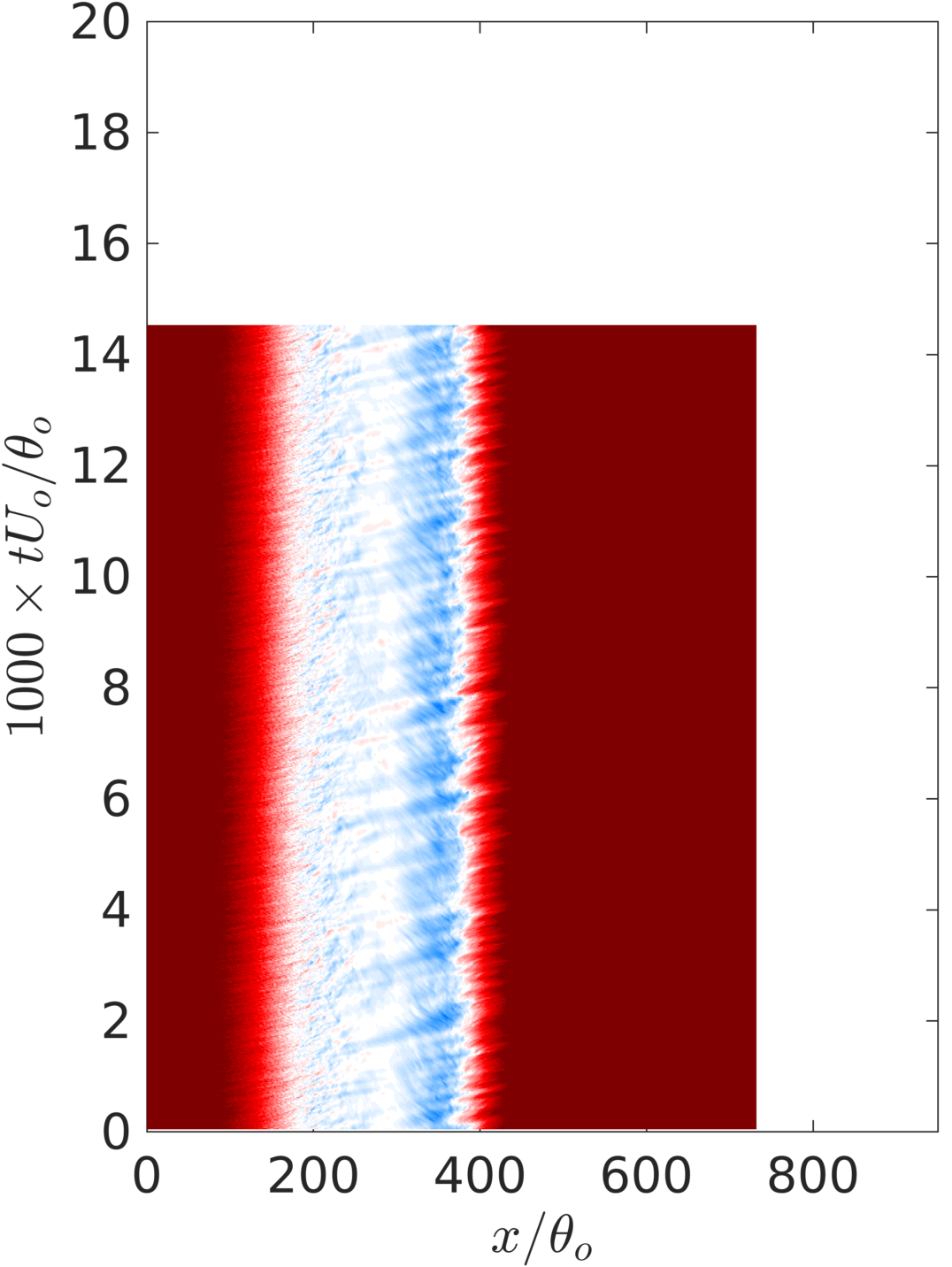}
  \includegraphics[height=0.55\textwidth]{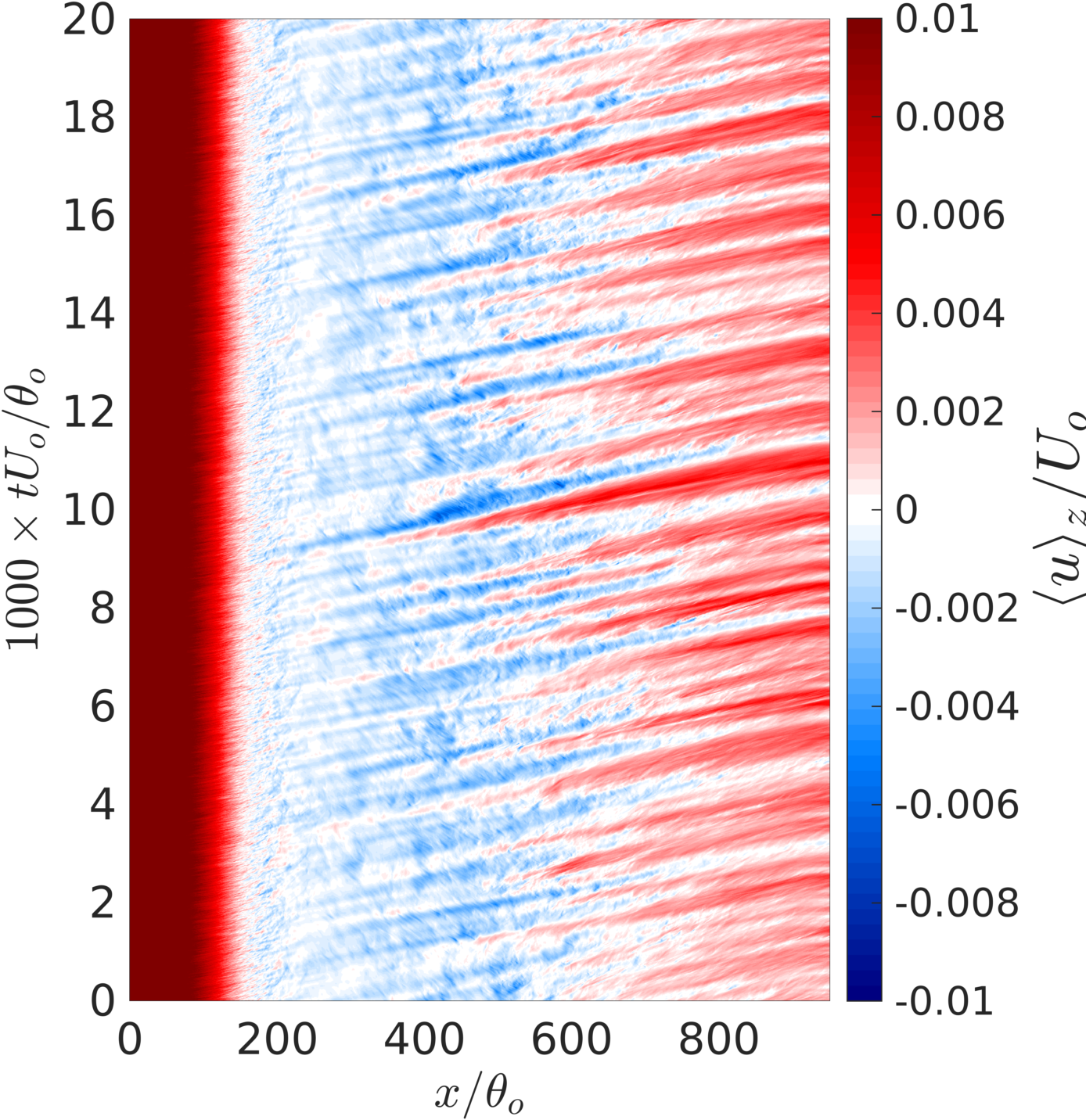}
%  \includegraphics[width=0.4\textwidth]{spatialtemporal_spectrum_new.png}
%  \caption{Left: spatial-temporal map of the spanwise-averaged streamwise velocity at the first grid point away from the bottom wall for the TSB-SO bubble. Right: pre-multiplied energy spectrum. Each plot is shifted upwards by 7$\times 10^{-7}$ units in the for clarity. Thin solid marks the high-frequency $f_h = 0.0025 U_o/\theta_o$, dashed represents the low-frequency $f_l = 0.001 U_o/\theta_o$, dotted-dash is $2f_l$. }
  \caption{Spatio-temporal map of the spanwise-averaged streamwise velocity at the first grid point away from the bottom wall for: left, TSB-SB bubble; right, TSB-SO bubble. Some region in the left subfigure is left blank intentionally to keep the two subfigure have the same axis range, because the TSB-SB case was integrated to $t = 14,300\,\theta_o/U_o$ with a smaller domain in $x$.}
  \label{fig:sptialtemporal}
\end{figure}

We begin by examining more closely the reversed flow in the vicinity of the wall for the TSB-SO bubble. The spatial-temporal map of the spanwise-averaged streamwise velocity at the first grid point away from the wall is plotted in Fig. \ref{fig:sptialtemporal}. For this separation bubble, besides the fluctuation of the incoming turbulent boundary layer at a very short time scale, a high frequency unsteadiness is featured as parallel stripes separated in time by about $TU_o/\theta_o \approx \mathcal{O}(400)$ in the map. Also, strong forward flow is observed to penetrate into the reversed flow region up to $x/\theta_o=400$ and this occurs at a time scale that is larger than that at the upstream region of the separation bubble.
The low frequency motion does not appear in the TSB-SB case, in which the reversed flow stripes exhibit the same temporal interval near the separation and reattachment regions.
%Note that high as well as low frequency unsteadiness observed in the TSB-SO cases are highly stochastic. 

\begin{figure}
\centering
  \includegraphics[width=0.4\textwidth]{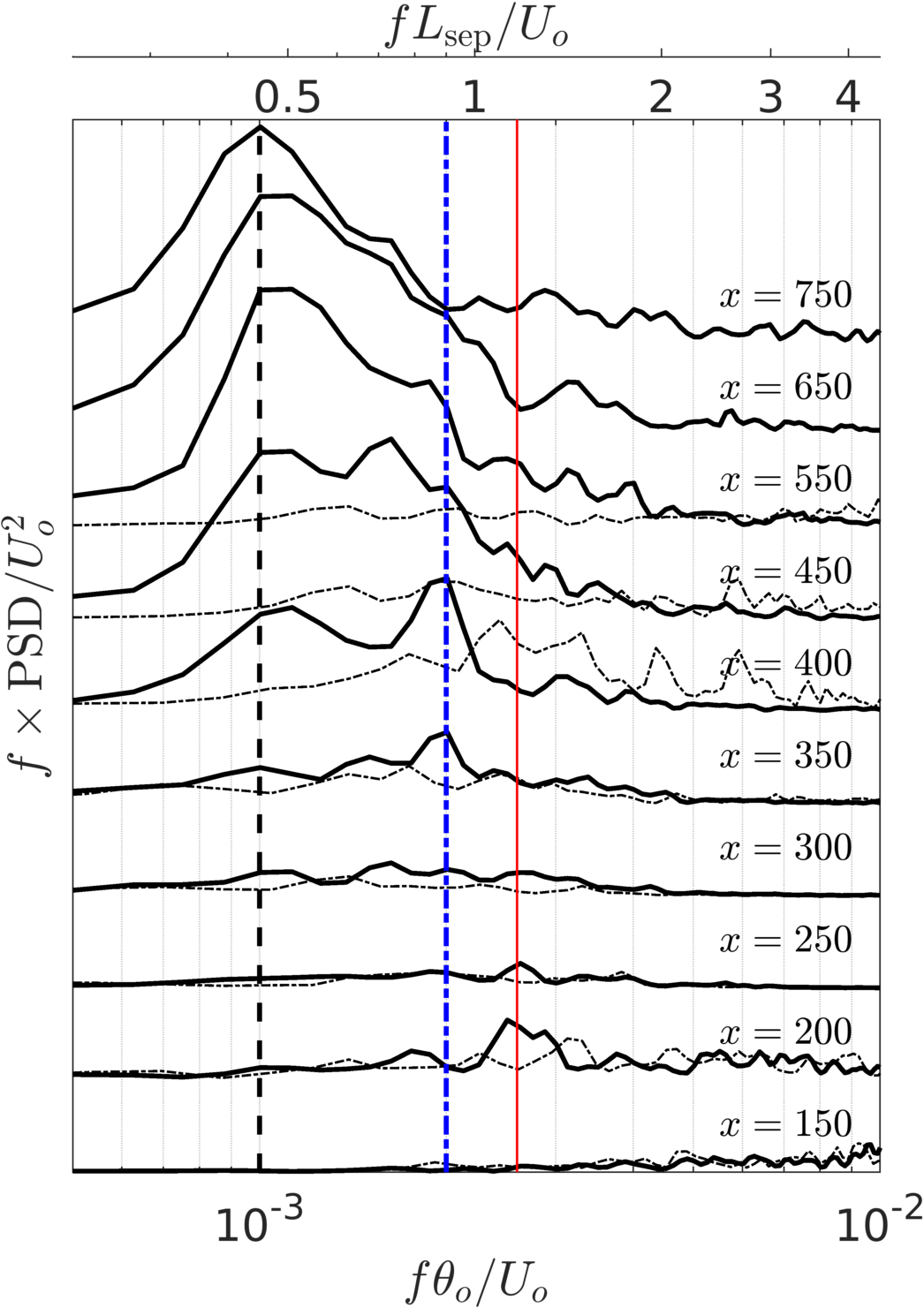}
  \caption{Pre-multiplied energy spectrum of the streamwise velocity showed in Fig. \ref{fig:sptialtemporal}. \revb{Dash-dotted, TSB-SB; solid, TSB-SO}. Each plot is shifted upwards by 7$\times 10^{-7}$ units in the for clarity. Thin vertical solid line marks the high-frequency $f_h = 0.0025\, U_o/\theta_o$, dashed vertical line represents the low-frequency $f_l = 0.001\, U_o/\theta_o$, dotted-dash vertical line is $f_m=0.002\,U_o/\theta_o$. }
  \label{fig:sptialtemporal_PSD}
\end{figure}

\reva{Note that the time scale showed here has no direct link to the ones described in Fig. \ref{fig:Axy}: Fig. \ref{fig:sptialtemporal}  represents the local (i.e., $x$) intermittency of the flow on the wall, describing the streamwise motion of a band-like, quasi-2D separated region. Fig.  \ref{fig:Axy}, on the other hand, is an integral quantity over the entire $x-y$ domain describing the total reversed flow area. When a discrete vorticity packet sheds off from the separating shear layer, for instance, it will show as unsteadiness in Fig. \ref{fig:sptialtemporal}  but not in Fig. \ref{fig:Axy} until the discrete vorticity packet decays and the reversed flow diminishes.}

Figure \ref{fig:sptialtemporal_PSD} plots the premultiplied power density spectra of the velocity shown in Fig. \ref{fig:sptialtemporal}. The data series is windowed and detrended and the analysis uses a total of eleven equal-length segments in time \revb{with 50\% overlap~\citep{NaM98b, Abe17}. Each segment contains 896 samples (stored every 100 time steps). The resulting resolved frequency range is $0\le f\theta_o/U_o \le 0.129$.} On the upstream end of the separation bubble (\textit{i.e.}, $x/\theta_o \approx 200$), a clear peak appears at a frequency of about  $f_h\theta_o/U_o \approx 0.0025$. At the other end of the separation bubble ($x/\theta_o>400$) the spectrum is dominated by a low-frequency corresponding to about $f_l\theta_o/U_o \approx 0.001$. The spectrum also indicates a peak corresponding to a frequency of $f_m\theta_o/U_o \approx 0.002$ in the region between 350-400 $\theta_o$. We refer to this as the ``mid-frequency" but as we will discuss later in the paper, this mode and the one at $f_l\theta_o/U_o \approx 0.0025$ seem to be driven by the same vortex rollup mechanism. Overall, the low and high frequencies show a clear separation in both scale ($f_h\theta_o/U_o \approx 0.0025$ versus 0.001) and location (high frequency for $x/\theta_o<$250 and low-frequency for $x/\theta_o>$250), with the $x/\theta_o$ from 350 to 400 appearing as a region of transition between these two distinct regions of the flow. \revc{The low-frequency frequency coincides with the subharmonic of the medium-frequency motion, which indicates that the former could be a result of vortex pairing.}

\begin{figure}
\centering
  \includegraphics[width=0.9\textwidth]{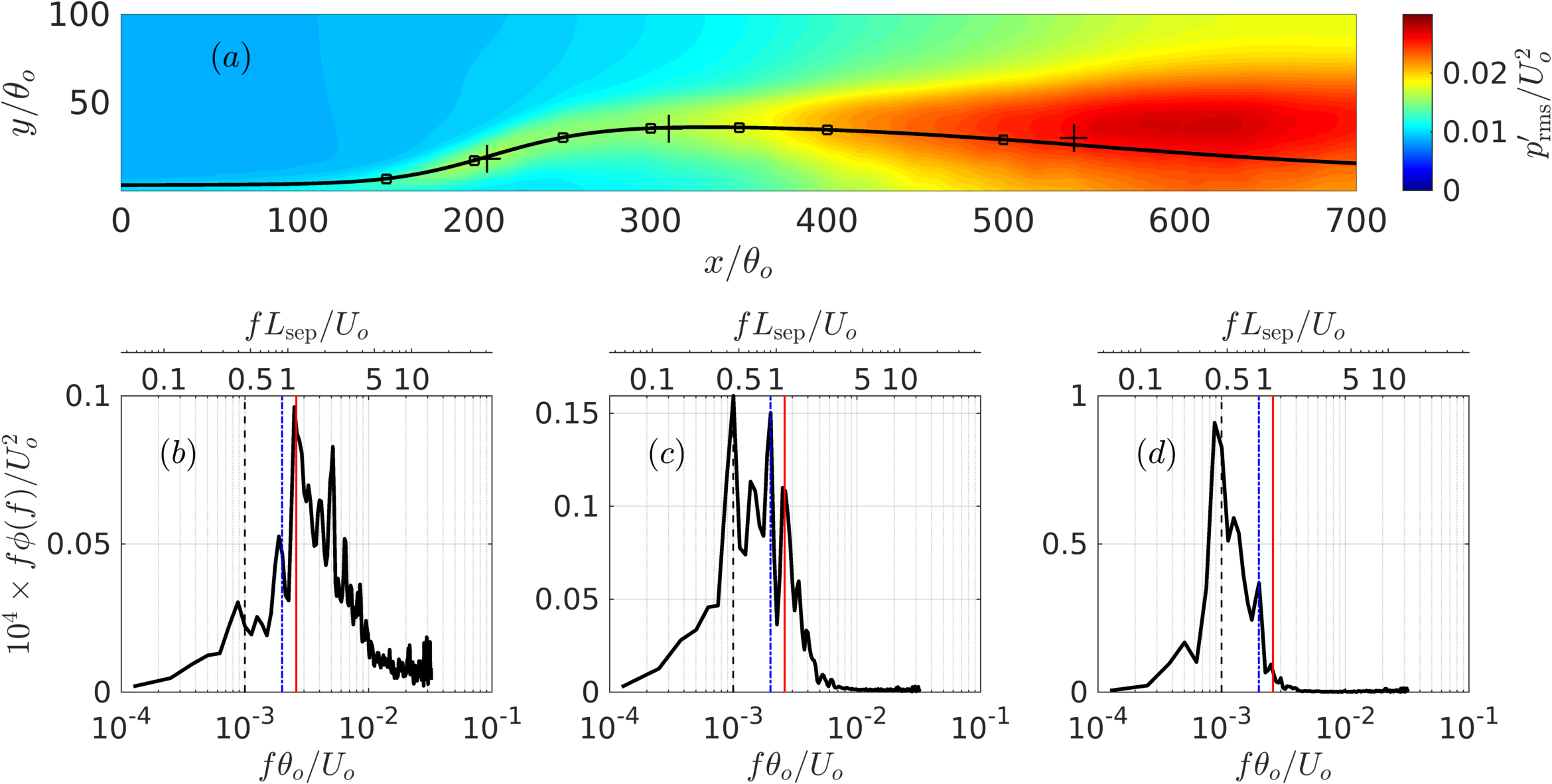}
  \caption{Pre-multiplied energy spectrum of the pressure fluctuation at several locations in the separated shear layer. (a) \revb{Contour map of $p^\prime_\text{rms}$}; the cross markers indicate the locations where the spectrum are obtained \revb{($x/\theta_0 = 207.3$ for (b), $310.3$ for (c) and $540.0$ for (d))}. The square markers are the locations where the two-point correlation is evaluated. \solid: selected mean streamline passing the high-$p^\prime_\text{rms}$ regions. (b-d)  pressure spectra. The thin solid vertical line marks the high-frequency $f_h = 0.0025 \,U_o/\theta_o$, the blue dashed vertical line represents the low-frequency $f_l = 0.001 \,U_o/\theta_o$, the dotted-dash vertical line marks $f_m= 0.002 \,U_o/\theta_o$.}
  \label{fig:p_PSD}
\end{figure}

\reva{The unsteadiness exhibited in the spanwise averaged field should not be interpreted as implying that the underlying mechanism is dominated by two-dimensional motions. Instead, it suggests that the spatial signature of the low frequency motion is visible in the the spanwise-averaged flow and this is indeed consistent with the notion of ``breathing" or expansion and contraction of the bubble.} 
While Fig. \ref{fig:sptialtemporal} corresponds to the data extracted very near the wall, similar frequency peaks and scale separation are also observed within the separated shear layer. Figure \ref{fig:p_PSD} (a) shows contours of \revb{the fluctuating pressure $p^\prime_\text{rms}$ and figure \ref{fig:p_PSD} (b-d) shows the pre-multiplied energy spectrum of pressure fluctuations at several locations within the high $p^\prime_\text{rms}$ regions}. \revb{Same as the velocity spectrum shown above,} the pressure data series are windowed and detrended with 50\% overlap, and the spectra is averaged over all the grid points in the spanwise direction at each location. The peaks of the spectrum agree well with those for the reversed flow on the wall. One difference is that the low-frequency motion already appear at $x=200\,\lo$ in the separated shear layer while it shows up much further downstream on the wall. This indicates that this low-frequency mode originates from the separated shear layer and impacts the near-wall flow after its effects spread downstream. Note that the ellipticity of the pressure does not seem to be the cause for this because such low-frequency motion is not observed in Fig. \ref{fig:sptialtemporal} and \ref{fig:sptialtemporal_PSD} near the separation point. 

%----------------
\reva{The Strouhal number defined as $St_{L_\text{sep}} = fL_\text{sep}/U_o$ are 1.125, 0.9 and 0.45 for $f_h$, $f_m$ and $f_l$, respectively in the TSB-SO case. The high-frequency mode also appears in the TSB-SB case but the Strouhal numbers differs by a factor of 2.4 due to the different $L_\text{sep}$. In previous studies using the suction-blowing configuration~\citep{NaM98a,Abe17}, a motion at $St\approx0.35$ has been reported as the `shedding mode'. In our TSB-SB case, this value is close to 0.4, which agrees reasonably well with the literature. The fact that the same phenomenon (i.e., KH vortex shedding) in the two configuration gives different $St_{L_\text{sep}}$ indicates that $L_\text{sep}$ is not the best parameter to characterize the unsteadiness of a separation bubble. 
} 
\revb{In the current study we associate the motion at frequency $f_l$ with the low-frequency motion even though its Strouhal number is higher than some previous studies, because this motion agrees phenomenologically with the ``breathing" phenomenon associated with the low-frequency motion in earlier studies of incompressible SBs~\citep{PauleyMR90,SpalartStrelets00,WeissTS15,TaifourW16}.}
%--------------------------

\begin{figure}
\centering
  \includegraphics[width=0.4\textwidth]{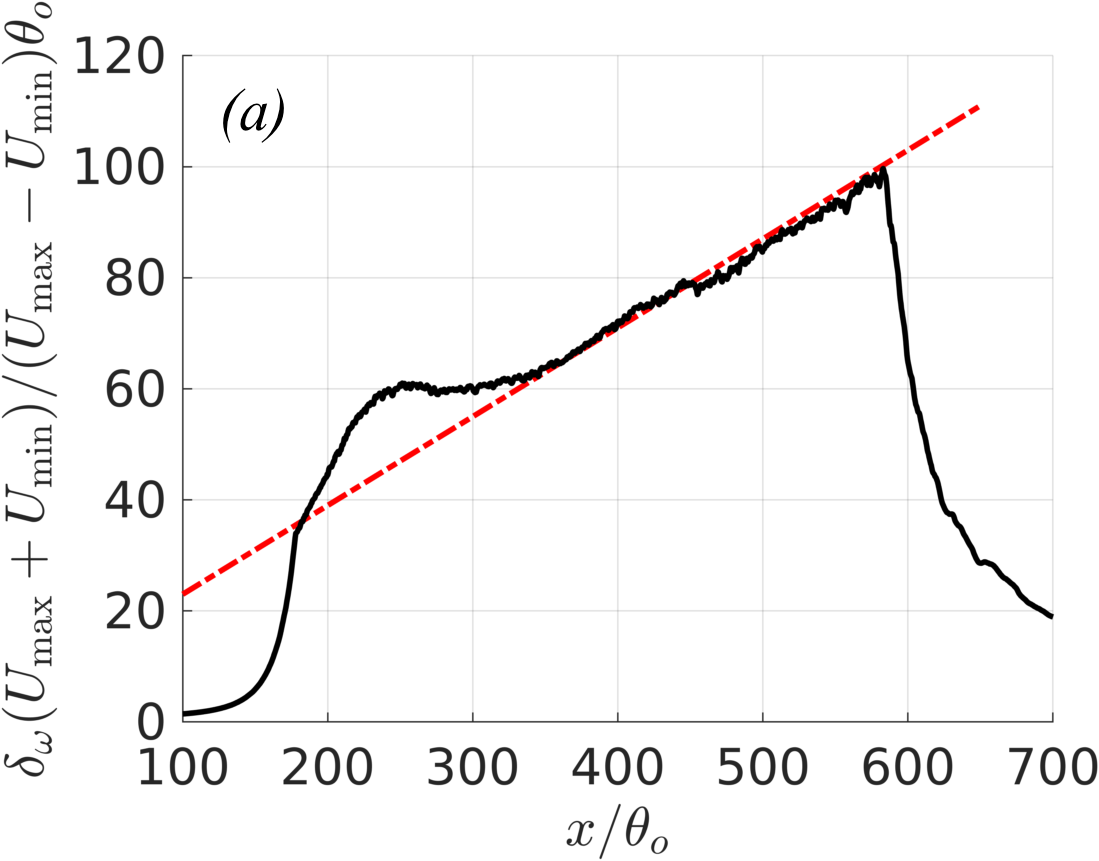}
  \includegraphics[width=0.4\textwidth]{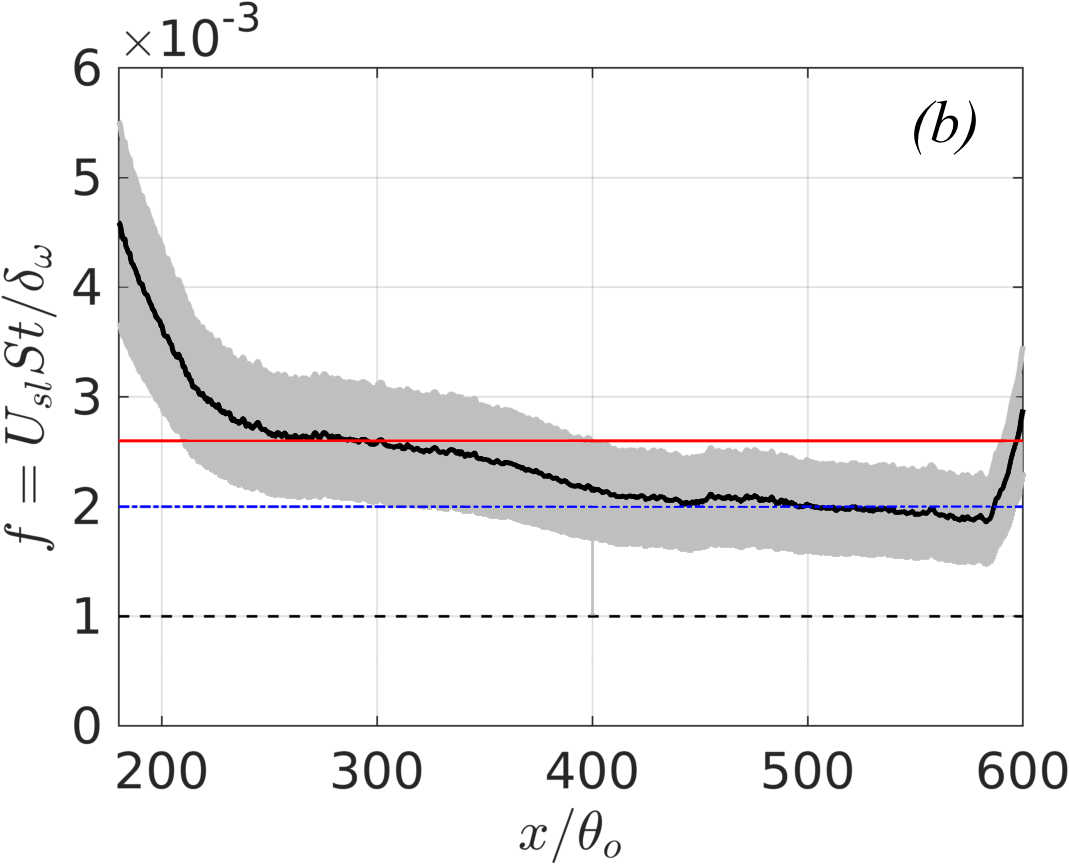}
  \caption{(a) Profile of the normalized vorticity thickness. \solid, TSB-SO, \dashdot, $d\delta_\omega/dx = C_{\delta_\omega} (U_\text{max}  - U_\text{min})/(U_\text{max}  + U_\text{min})$, $C_{\delta_\omega}=0.16$. (b) Dominant frequency of mixing layer predicted by canonical mixing layer relationship. \solid, $St_{\delta_\omega} = 0.25$; gray hatched region, $St_{\delta_\omega} \in [0.2, \,0.3]$. The thin solid horizontal line marks the high-frequency $f_h = 0.0025\, U_o/\theta_o$, the dotted-dash horizontal line marks $f_m= 0.002 \,U_o/\theta_o$, and the dashed horizontal line represents the low-frequency $f_l = 0.001\,U_o/\theta_o$, .}
  \label{fig:domega}
\end{figure}

%\label{sec:ml}
\revb{Previous studies have shown that the separated shear layer is similar to a canonical plane turbulent mixing layer (e.g., velocity and Reynolds stress scale well by the mixing layer scaling) and the high-frequency motion may be generated by Kelvin-Helmholtz instability (e.g., large-scale spanwise structures)~\citep{NaM98a,SpalartStrelets00,RistM02,MarxenH10,AbeMMS12,Abe17}}. To examine the degree of similarity to a mixing layer, we have examined the profile of the vorticity thickness $\delta_\omega$ defined as
\begin{equation}
\delta_\omega = (U_\text{max}-U_\text{min})/(\partial U/\partial y)_\text{max}
\end{equation}
where $U_\text{max}$ and $U_\text{min}$ are the maximum and minimum time- and spanwise-averaged streamwise velocity in the two sides of the separated shear layer. We find that (see Fig. \ref{fig:domega} (a)) $\delta_\omega$ exhibits a linear growth with $x$ after an initial transition at the beginning of the separation. The growth rate also agrees very well with the one found in previous studies on plane mixing layers\revb{~\citep{LiepmannL47,BrownR74,PantanoS02,Abe17}}. 

It has been widely reported in previous studies that the dominant frequency in a turbulent mixing layer corresponds to a Strouhal number $St_{\delta_\omega} = f\delta_\omega/U_{sl} \approx 0.2-0.3$ ($U_{sl} = (U_\text{max}+U_\text{min})/2$). Figure \ref{fig:domega} \textit{(b)} plots the frequency predicted by this relationship in the current simulation where we have used the local $\delta_\omega$. It can be seen that with $St_{\delta_\omega}=0.25$ the scaling predicts a frequency $f\theta_o/U_o$ ranging from about 0.004 to 0.002 in the $200<x/\theta_o<250$ region, which is very much inline with the  two observed high frequencies of $f_m\theta_o/U_o=0.002$ and $f_h\theta_o/U_o = 0.0025$ in this region of the bubble. In particular, the profile shows one plateau upstream of $x/\theta_o=350$ and another downstream. It agrees very well with the locations where the spectral peaks appear in Fig. \ref{fig:sptialtemporal_PSD}. This provides strong support for the notion that the mechanism of high- and medium-frequency unsteadiness is the KH instability  of  the  mixing  layer.  Moreover,  the  plot  shows  that  despite  a  decrease in  this  predicted  frequency  with  downstream  distance,  it  never  reaches  the  observed low frequency of $f_l\theta_o/U_o$= 0.001. This suggests that some mechanism other than KH instability drives the low-frequency unsteadiness in the rear part of the separation bubble.

\begin{figure}
\centering
  \includegraphics[width=0.8\textwidth]{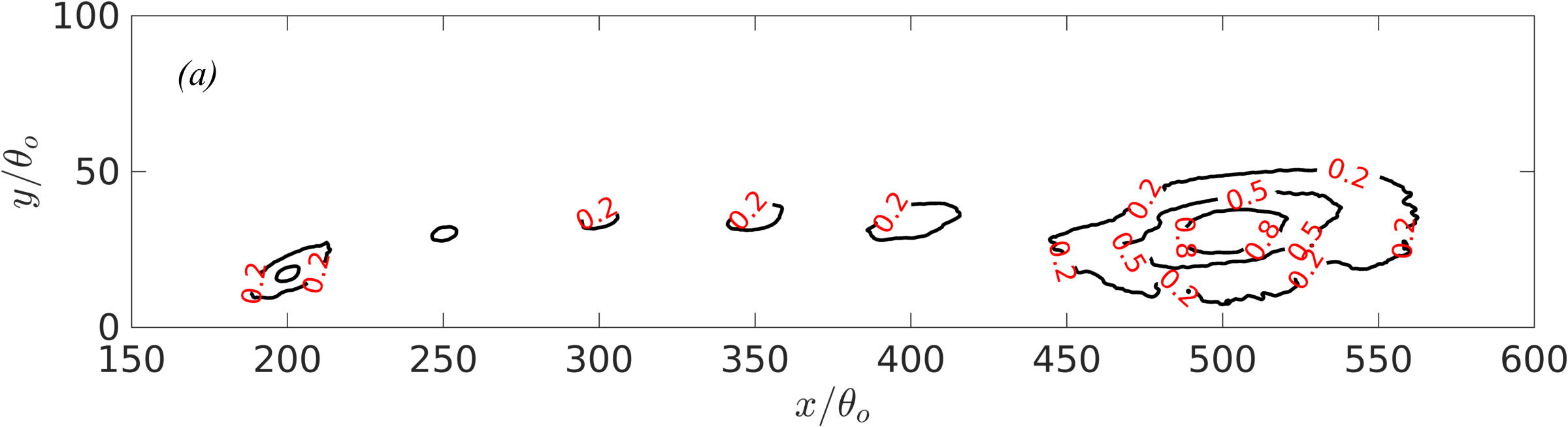}
  \includegraphics[width=0.8\textwidth]{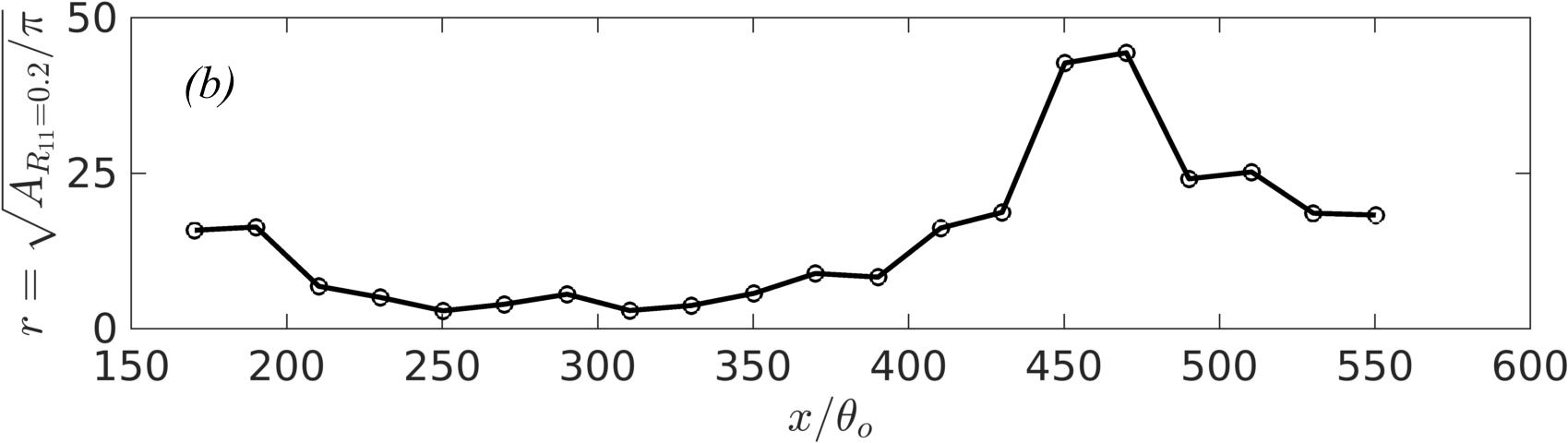}
  \caption{(a) Contours of two-point correlation coefficient of streamwise velocity fluctuations with levels of correlation indicated by text. The reference locations are selected along a mean streamline passing the high-$p^\prime_\text{rms}$ regions and are indicated by square markers in Fig. \ref{fig:p_PSD} (a). The contours for different reference locations are superposed. (b) the nominal radius of the roller vortex determined from the area enclosed by the $R_{11} = 0.2$ contourline, \textit{i.e.}, $r/\theta_o=\sqrt{A/\pi}$. }
  \label{fig:2pcoef}
\end{figure}
 Because the vortices generated by KH instability are spanwise rollers, we have examined the length scale of the flow in the separated shear layer, using the two-point autocorrelation coefficient in the $x-y$ plane
\begin{equation}
R_{11}({\bf x}_\text{ref},{\bf x}_\mathrm{ref} + \Delta {\bf x}) = \frac{\langle u^\prime({\bf x}_\mathrm{ref})u^\prime({\bf x}_\mathrm{ref}+\Delta {\bf x})\rangle}{\langle u^\prime({\bf x}_\mathrm{ref})u^\prime({\bf x}_\mathrm{ref}) \rangle}\end{equation}
where $\Delta {\bf x}$ is the spatial separation between the reference location and the other locations in the computational domain.
Several reference locations are chosen along a mean streamline passing through the high-$p^\prime_\text{rms}$ regions and the correlation is calculated between each of them and all the other location in the flow fields. The contours of the correlation coefficients are shown in Fig. \ref{fig:2pcoef} where the $R_{11}$ contour lines for selected reference location are presented on the same plot. If we define non-negligible correlation as $R_{11} \geq 0.2$, it can be seen (Fig. \ref{fig:2pcoef} (b)) that the spanwise roller starts to generate near $x/\theta_o=250$ and gradually grows in size. Between $x/\theta_o=$ 400 and 500, the vortex length-scale increases abruptly, suggesting the formation of large-scale vortex structures, possibly associated with the merging of the spanwise rollers. 

\begin{figure}
\centering
  \includegraphics[width=0.95\textwidth]{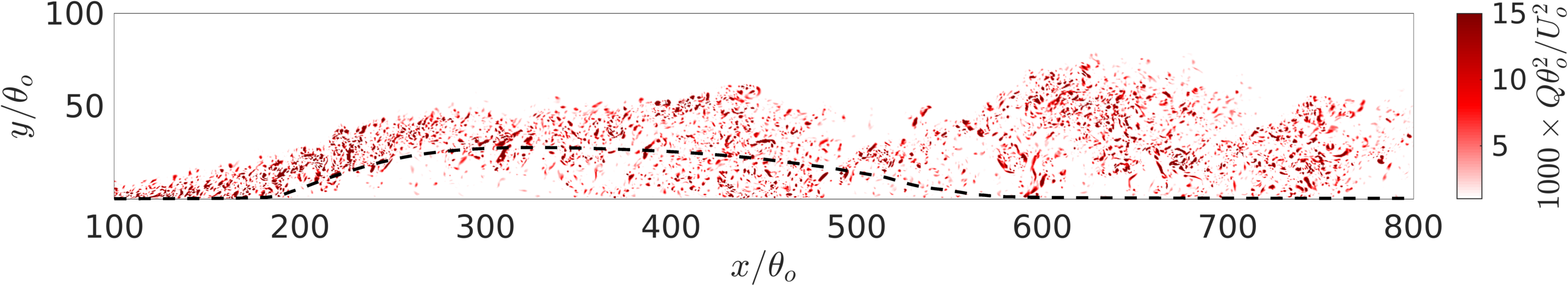}
  \caption{Contours of instantaneous secondary invariant of the velocity gradient tensor in the $x-y$ plane $z=L_z/2$ at  $tU_o/\theta_o=16,950$. Only the positive $Q$ are shown for clarity. The mean separating streamline is showed as the dashed line for reference.}
  \label{fig:contourQ}
\end{figure}

Contours of the second invariant of the velocity gradient tensor at one selected time (corresponding to the time-instance shown in Fig. \ref{fig:Qiso} (b)) are shown in Fig. \ref{fig:contourQ} and in the turbulent shear layer, the spanwise rollers consist of small-scale eddies. A large-scale conglomerations of vortices is formed near $x/\theta_o=450$ at the time shown and this seems to be inline with the notion that the large scales at this downstream location may be caused by a vortex merging process.  
  
It is however not clear whether this vortex merging is the mechanism underlying the low-frequency unsteadiness or whether the latter arises due to some  other distinct mechanism. In particular, there are a number of possible candidate mechanisms for the low-frequency unsteadiness of the separation bubble. These include:
\begin{enumerate}[align=left,leftmargin=3em,itemindent=0pt,labelsep=0pt,labelwidth=2em]
\item Imbalance of entrainment from the recirculation region by the shear layer and the reinjection of fluid near the reattachment point~\citep{EatonJ82}; this makes the separating shear layer flap towards the wall and significantly slows down its convection downstream.
\item Shear layer switches from a convective instability to a global instability which leads to local amplification. In laminar flows, researchers have found that global instability occurs when the mixing layer is formed by counter streams and when $R_u=(U_\text{max}-U_\text{min})/(U_\text{max}+U_\text{min})$ exceeds 1.315~\citep{HuerreM85, StrykowskiN91}, or the peak reversed flow amplitude exceeds 20\% of the freestream velocity~\citep{HammondR98,RistM02}. In the current flow the maximum value of $R_u$ is 1.25 and the peak reserved flow magnitude is 0.08 $U_o$. Thus, this is not likely to be the mechanism. 
\item Some mechanism unrelated to the KH vortices, which breaks down the sheet of the spanwise vortices. 
\end{enumerate}

In general, it is not straightforward to separate cause from effect in a fully saturated, highly turbulent flow with a large range of temporal and spatial scales. In the current study, we employ dynamic mode decomposition (DMD)~\citep{Schmid10,ChenTR12} to examine the flow field and to extract features/modes of the flow that correspond to the dominant time-scales in the flows. The topology of these extracted modes is then used as the basis to further investigate the underlying mechanism for the low-frequency ``breathing" mode. 

\begin{figure}
\centering
  \includegraphics[width=0.8\textwidth]{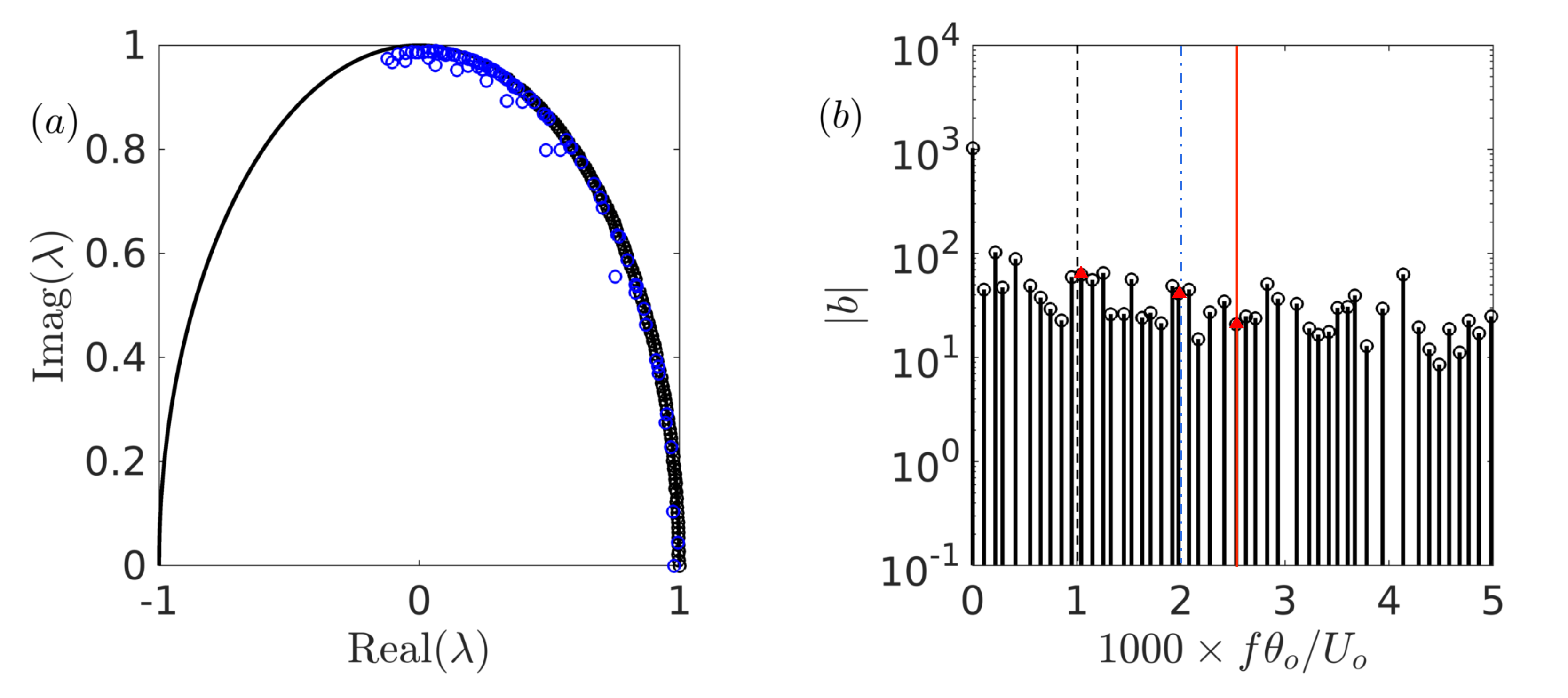}
  \caption{(a) Eigenvalues of the DMD modes; (b) Spectrum of the DMD modes; the range of $x$-axis is limited to the frequency range of concern. Modes whose $\vert \lambda \vert<0.995$ are showed by blue marker in (a) and not included in (b). The three thin vertical lines refer to Fig. \ref{fig:p_PSD}. The triangular marker shows the selected DMD modes closest to the three target frequencies.}
\label{fig:dmd_spectrum}
\end{figure}

\subsection{Dynamic Mode Decomposition based Analysis of the Three-dimensional Flow Field}
The basic idea of DMD analysis is to extract a low-dimensional description of a linear transformation that maps any snapshot of data from a dynamical system into the subsequent snapshot. The eigenvalue decomposition of this linear transformation then provides frequency information as well as the corresponding spatial structures. Furthermore, a projection of the snapshots onto these structures gives the amplitude (contribution) information~\citep{Schmid10,Sayadietal14}. Eventually, the snapshots can be approximated using the DMD modes as 
\begin{equation}
\Phi_r(\mathbf{x},t) = \sum_{k=1}^r b_k \psi_k(\mathbf{x}) \exp(\omega_k t), \; k =1,2,...,r
\end{equation}
in which $\Phi_r$, $\psi_k$ and $b_k$, are the field reconstructed by the $r$ DMD modes, the $k$-th spatial DMD mode (shape of the mode), and the magnitude of the $k$-th DMD mode, respectively. The term $\exp(\omega t)= \lambda^{t/\Delta t}$, where $\lambda$ is the eigenvalue of the modes, describes the growth/decay and oscillation of the mode. 

In this study, our objective of using DMD is to extract the modes corresponding to the key frequencies identified in the earlier part of the paper and to examine the topology of the modal reconstructions based on these modes as a means for identifying potential generation mechanisms. Discrete Fourier transform (DFT) based techniques such as notch filtering could be used for the purpose of modal reconstruction but these techniques requires the data series to cover an integer number of the corresponding periods, and suffers from spectral leakage~\citep{Sayadietal14}. Proper-orthogonal decomposition (POD) analysis is another decomposition approach that could be used but POD modes may contain multiple frequencies and therefore are not suitable for our analysis.

\begin{figure}
\centering
  \includegraphics[width=0.9\textwidth]{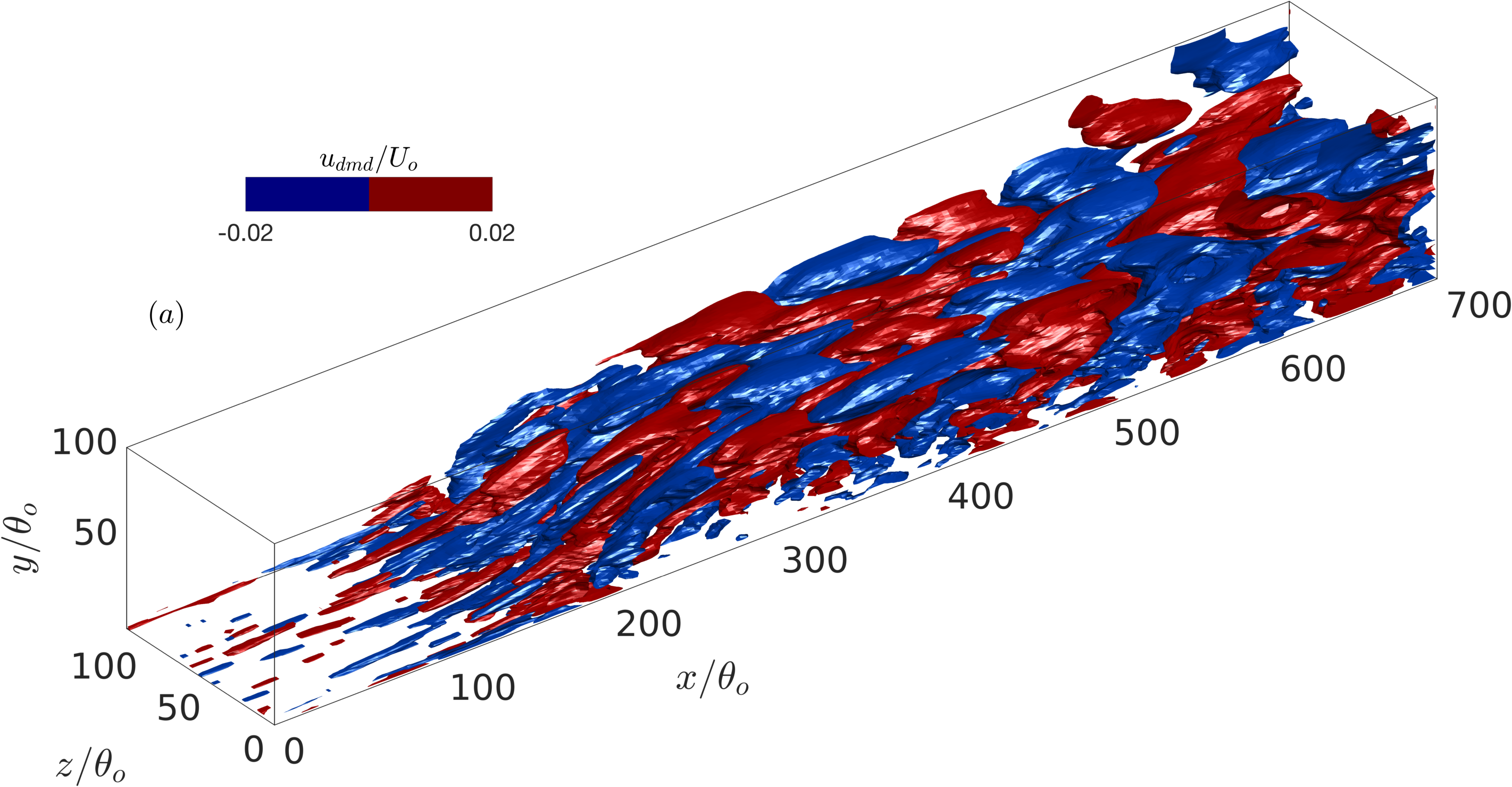}
  \includegraphics[width=0.9\textwidth]{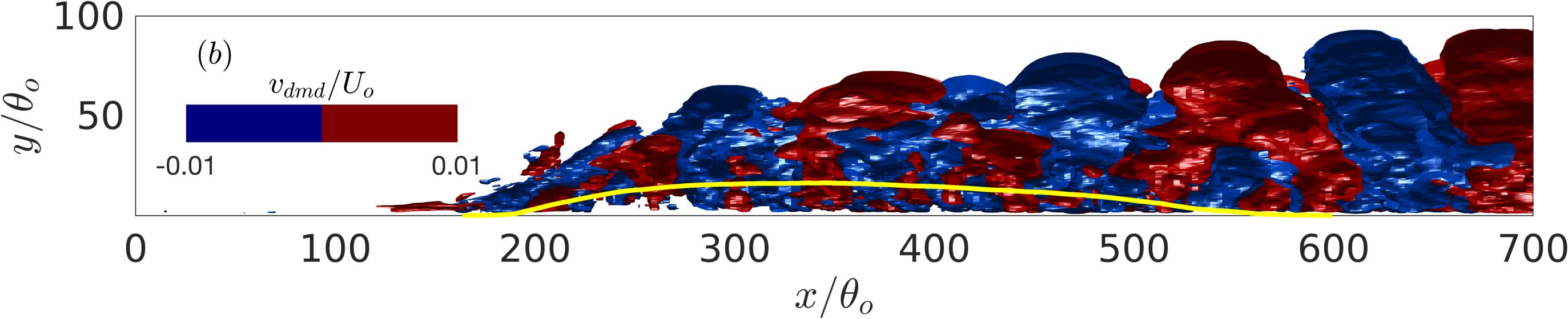} \includegraphics[width=0.9\textwidth]{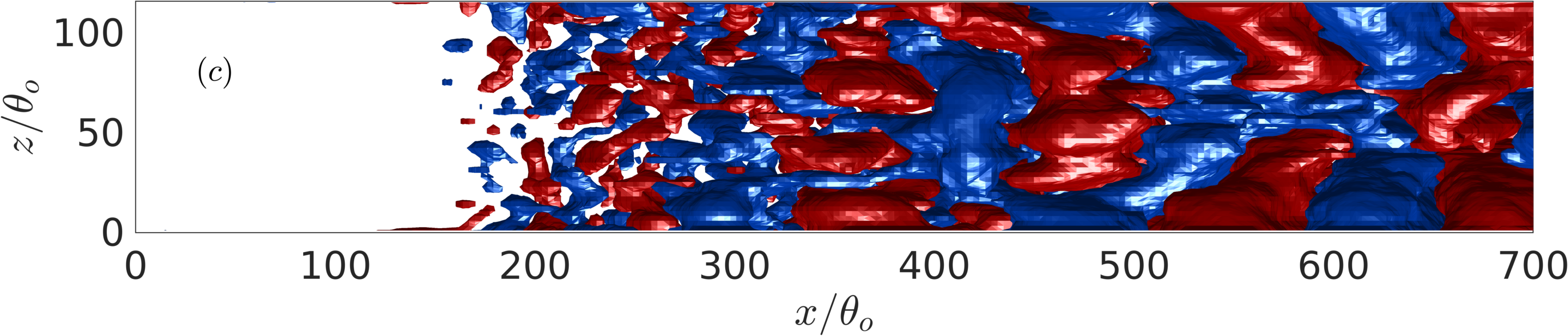} 
  \caption{Isosurfaces of the real part of the high-frequency DMD mode $f_h\theta_o/U_o = 2.49\times10^{-3}$, \revall{$St_{L_\text{sep}}=1.125$}. (a), $u_\text{dmd}/U_o=\pm0.02$;  (b, c), $v_\text{dmd}/U_o=\pm 0.01$ top and side view respectively. \solid $U=0$.}
  \label{fig:udmd_fh}
\end{figure}

\begin{figure}
\centering
  \includegraphics[width=0.9\textwidth]{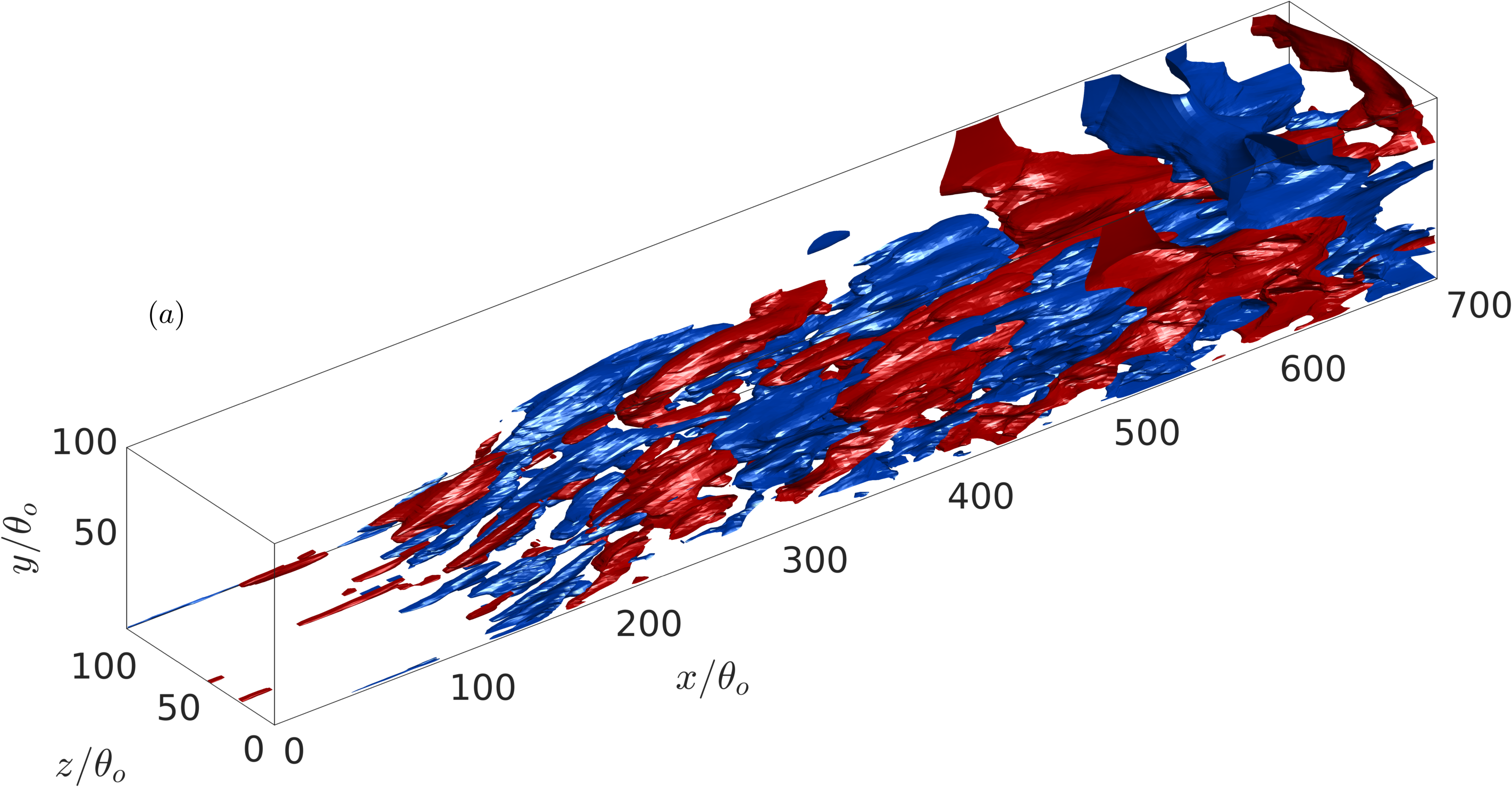}
  \includegraphics[width=0.9\textwidth]{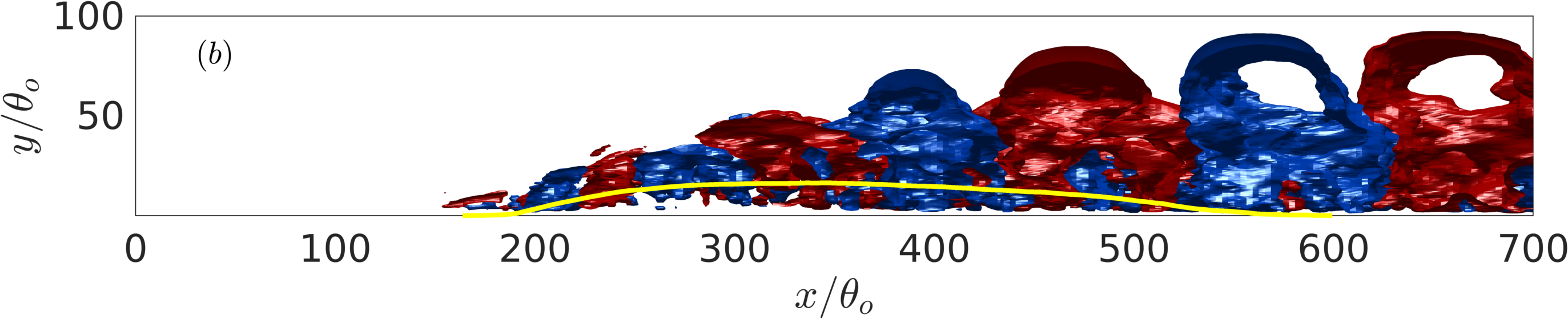} \includegraphics[width=0.9\textwidth]{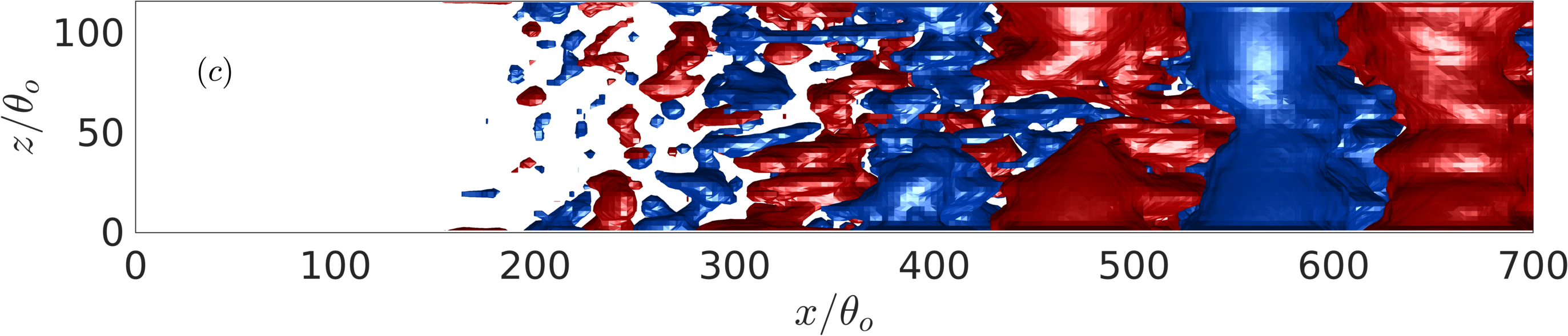} 
  \caption{Isosurfaces of the real part of the mid-frequency DMD mode $f_m\theta_o/U_o = 1.97\times10^{-3}$, \revall{$St_{L_\text{sep}}=0.90$}. Color style refers to Fig. \ref{fig:udmd_fh}. }   
  \label{fig:udmd_fh2}
\end{figure}

The total sample for the DMD analysis extends over a period of 11,230 $\theta_o/U_o$, which corresponds roughly to 11.2 periods of the observed low frequency. Snapshots are extracted at every 400 time steps; this corresponds to a sampling frequency of $0.0642 \,U_o/\theta_o$ and results in the extraction of a total of 721 snapshots. Each sample consists of the three velocity components $u$, $v$ and $w$. In order to capture the key regions of the separated and reattaching flow, the spatial region over which the data is extracted covers $0 \leqslant x/\theta_o \leq 700$ and $0 \leqslant y/\theta_o \leq 100$ . 

In order to reduce the overall size of the dataset used in the DMD analysis to manageable levels, we employ a spatial subsampling where we first filter the velocity components by a box-filter with width of 3 $\theta_o$, and then extract every eighth grid points in the $x$ and $z$ directions and every other point in $y$. This procedure leads to a total of 2.35 million samples per snapshot per observable, which is about 0.8\% of the total mesh nodes in the sample region. Finally, to avoid over-fitting the complex dynamics of the fully turbulent field, instead of using the velocity data directly, we employ a 361 mode POD projection (one real mean mode and 180 complex conjugate complex pairs) which serves as a spatio-temporal filter. This projection retains 98\% of the total kinetic energy, and 94\% of the turbulent kinetic energy of the flow. 

The eigenvalues and (discrete) spectrum of the DMD modes are shown in Fig. \ref{fig:dmd_spectrum}. Since the flow is in a statistically saturated state, the eigenvalues lie mostly on or slightly inside the unit circle on the complex plane. Furthermore, due to the broadband nature of the flow, the spectrum exhibits a relatively continuous distribution of modes without any distinct and isolated peaks. This indicates that a significant number of modes are required to accurately reconstruct the dynamics of this flow.
In the following, we focus on three
modes whose frequencies are closest to those identified as the high ($f\theta_o/U_o = 0.0025$), medium ($f\theta_o/U_o=0.002$), and low ($f\theta_o/U_o = 0.001$)  frequencies in the previous sections, \textit{i.e.}, $f_h\theta_o/U_o = 2.49\times10^{-3}$, $f_m\theta_o/U_o=1.97\times10^{-3}$ and $f_l\theta_o/U_o = 1.03\times10^{-4}$.

Figure \ref{fig:udmd_fh} and \ref{fig:udmd_fh2} shows DMD modes corresponding to the high- and mid-frequency spectral peaks. The key characteristic of these two modes that is visible from these plots is the highly regular arrangement of alternating structures separated in the streamwise direction. These structures also exhibit a considerable degree of coherence in the spanwise direction (Fig. \ref{fig:udmd_fh} (c), \ref{fig:udmd_fh2} (c)). \revc{The high-frequency mode appears similar to the $\lambda$-vortex structure formed by the oblique modes in transitional boundary layer~\citep{SchmidH12}}. Both these features correspond well to the notion that these two modes are associated with the KH instability and the modal structure corresponds primarily with spanwise vortex rollers.  
%The spanwise coherence of the structures in this high-frequency mode is clearly seen in Fig. \ref{fig:udmd_fh} (b, c). 
%The high-frequency mode is more three-dimensional and exhibits staggered positive-negative distribution in the streamwise direction, representing the spanwise coherence of the vorticity generated by the KH instability.
%Significant perturbations are observed after the flow is detached from the wall. .
% The amplification of the high-frequency perturbation starts from the separation point and occurs in the middle of the separated shear layer. This again proves that the mechanism generates this mode is the KH instability that is most amplified at the inflection point of the velocity profile within the mixing layer.
 
 \begin{figure}
\centering
  \includegraphics[width=0.9\textwidth]{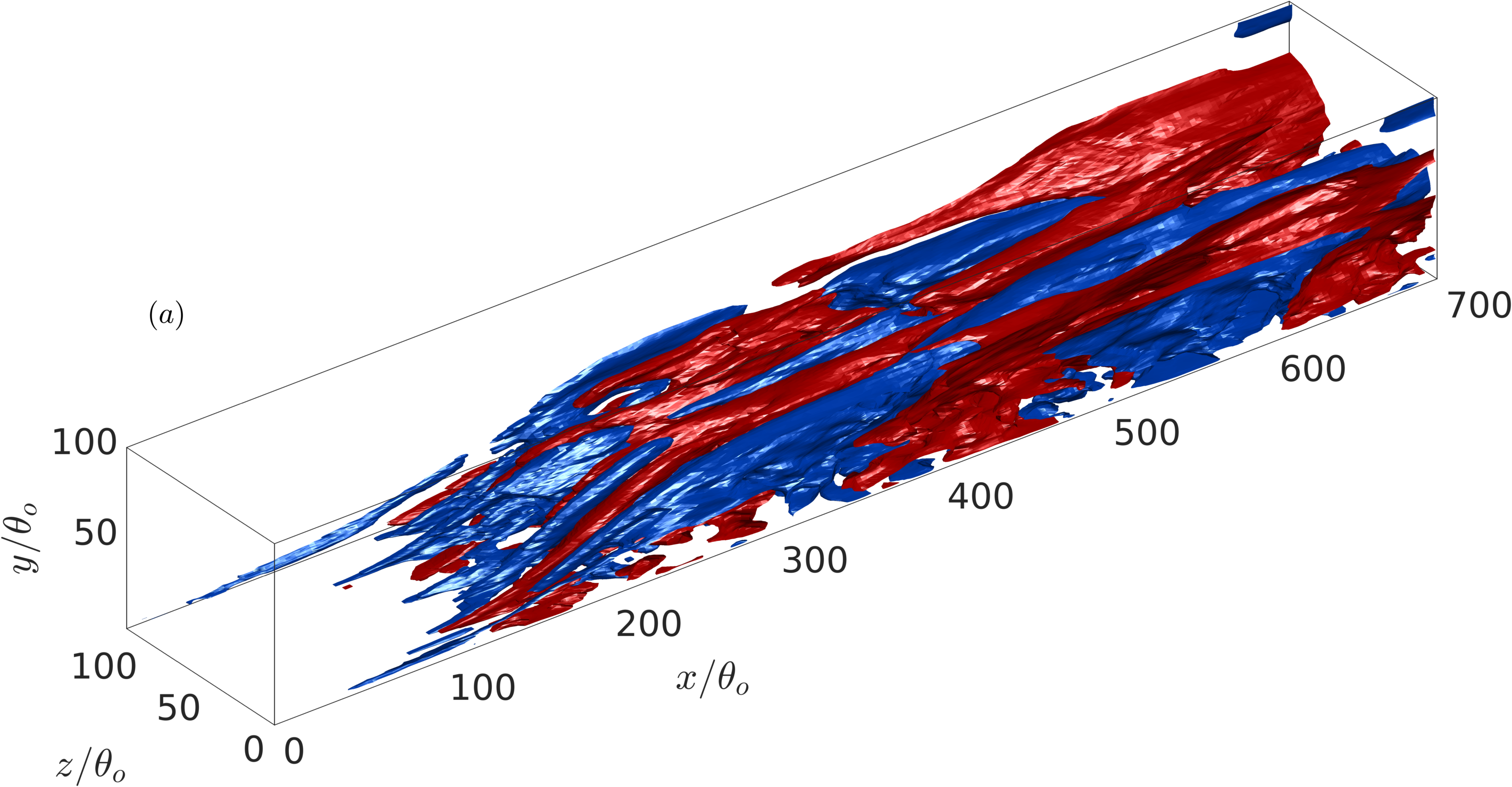}
  \includegraphics[width=0.9\textwidth]{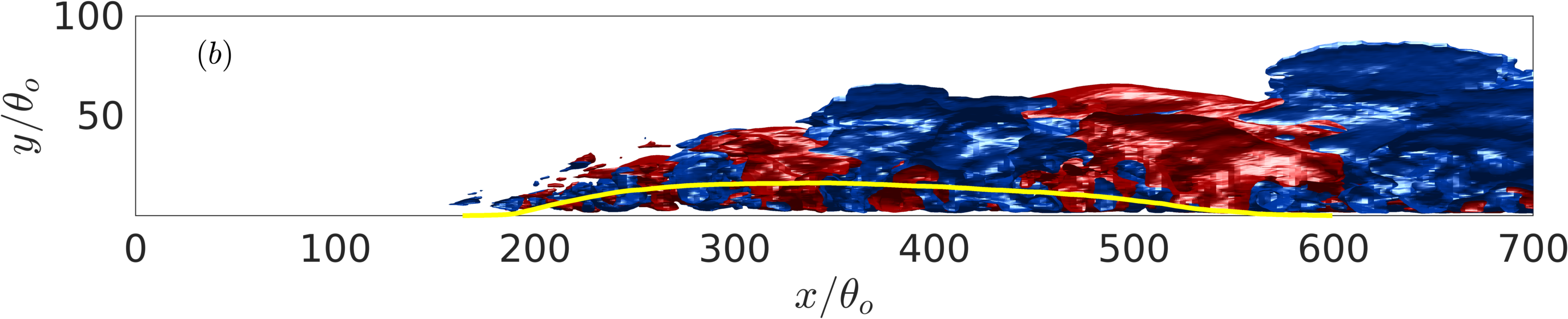} \includegraphics[width=0.9\textwidth]{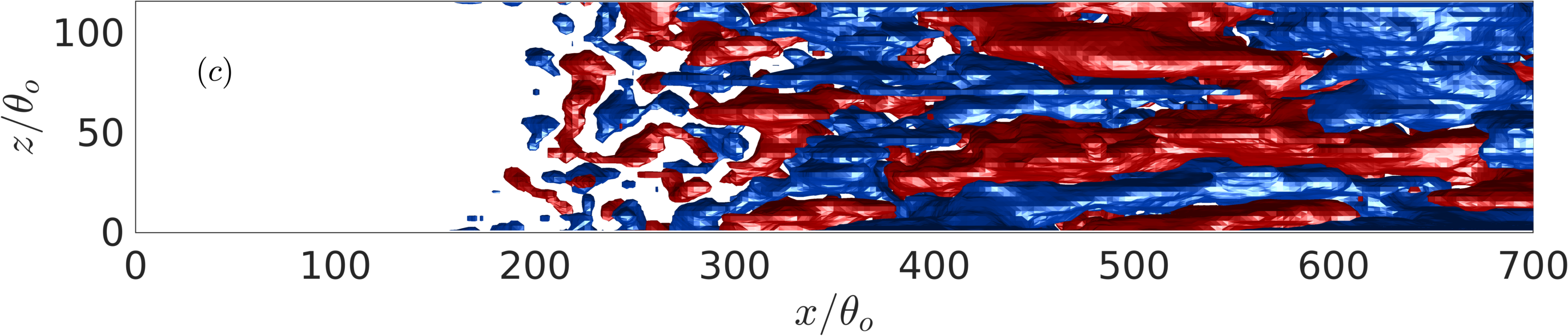} 
  \caption{Isosurfaces of the real part of the low-frequency DMD modes $f_l\theta_o/U_o = 1.03\times10^{-3}$, \revall{$St_{L_\text{sep}}=0.45$}.  Line style refer to Fig. \ref{fig:udmd_fh}}   
  \label{fig:udmd_fl}
\end{figure}
The topology of the low-frequency mode (shown in Fig. \ref{fig:udmd_fl}) is quite different from that of the high-frequency modes discussed above and appears to be dominated by highly elongated streamwise structures. For example, the structure located near $z=0$ starts from $x/\theta_o = 40$ and extends to $x/\theta_o = 410$; another one generated around  $[x, z]/\theta_o = [100, 86]$ extends to $x/\theta_o = 270$ (Fig. \ref{fig:udmd_fl} (a)). Note that the simulation is periodic in the spanwise direction so that the structure at $z=L_z$ connects with the one at $z=0$.   
Further downstream in the region where the freestream APG vanishes, the streamwise structures connect and form larger structures. However, their two-dimensionality along the streamwise direction seems to be preserved. 
When compared with the high-frequency modes, the low-frequency mode has a much larger wavelength in the streamwise direction (Fig. \ref{fig:udmd_fl} (a)). The size of the structures  grows abruptly near $x/\theta_o=400$ (Fig. \ref{fig:udmd_fl} \textit{(b,c)}).

\begin{figure}
\centering
  \includegraphics[width=0.98\textwidth]{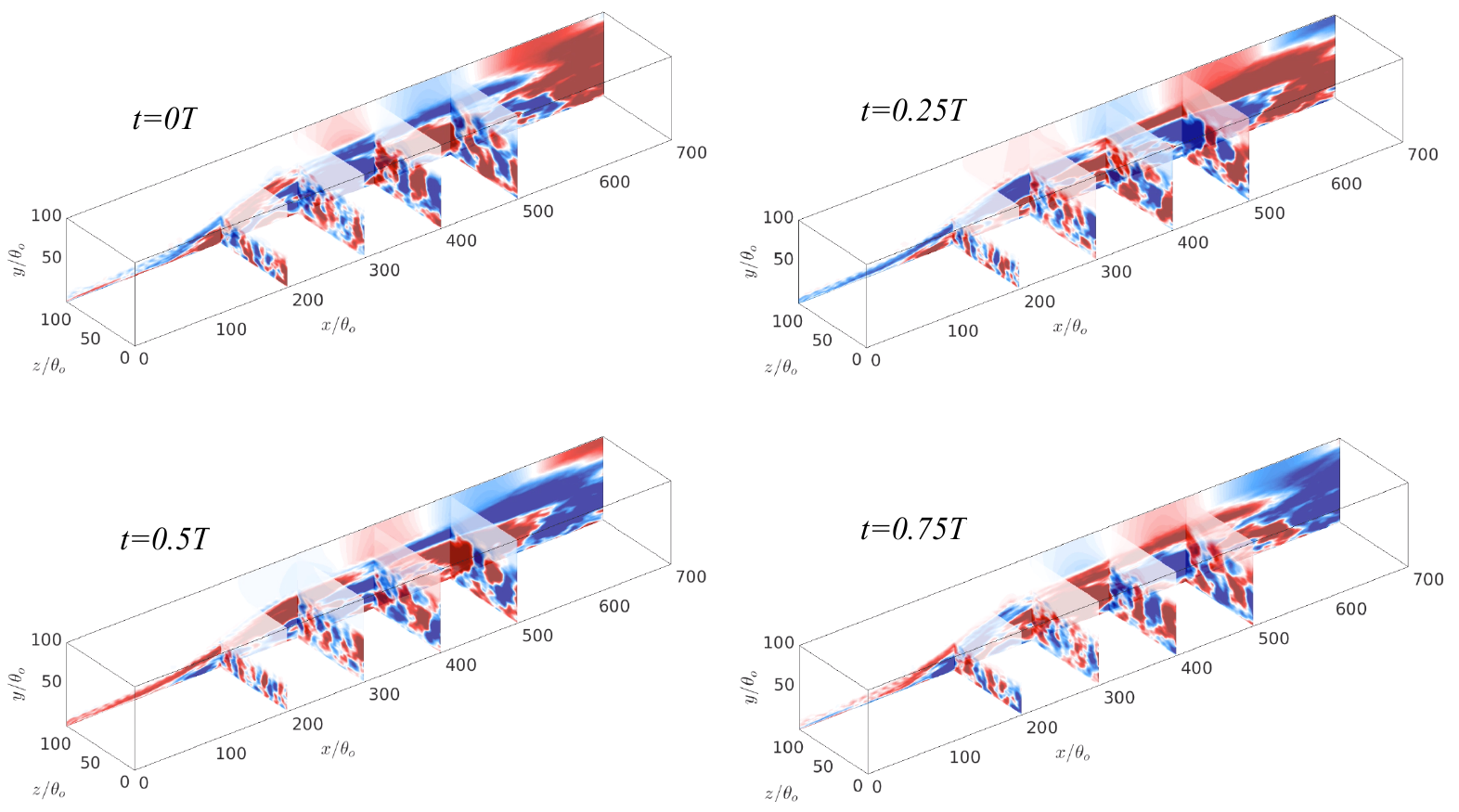}
  \caption{Contour of the streamwise velocity of the low-frequency DMD mode (refer to Fig. \ref{fig:udmd_fl}) at four equally-spaced phases in one cycle (i.e., period of $T$).}
  \label{fig:fl_dmd_phase}
\end{figure}

\revc{The side view of this mode shows that there is also a significant spanwise oriented signature in this mode. Furthermore, even structures that show a streamwise oriented topology can ``shed'' in a spanwise homogeneous manner resulting a noticeable spanwise-averaged signal, and that is what is likely what is happening in this flow: Fig. \ref{fig:fl_dmd_phase} shows the streamwise velocity component for the low-frequency DMD mode at four equally-spaced phases in one cycle. The shedding of structures can be observed in the $x-y$ slice contour, while the streamwise coherence persists in the $x-z$ slices.}

\subsection{G\"{o}rtler Instability as a Mechanism for the Low-frequency Mode}
The topology of the low-frequency mode suggests that it could be related to G\"{o}rtler vortices, which appear as streamwise elongated structures in the boundary layers over concave walls.  These G\"{o}rtler vortices are generated by a centrifugal instability that destablizes the flow when the direction of the wall-normal velocity gradient is opposite to the centrifugal force associated with the curvature of the streamlines~\citep{Gortler54}, and previous studies on separated flows have suggested the presence of G\"{o}rtler-type vortices~\citep{SettlesFB79, LoginovAZ06, PriebeTRM16}. Numerous previous studies on G\"{o}rtler vortices have proposed a variety of threshold criteria for G\"{o}rtler instability as well as the characteristic wavelength of the generated structures. In general, it is widely accepted that G\"{o}rtler instability appears when the G\"{o}rtler number, defined as
\begin{equation}
G_T = \frac{U_e\theta}{\nu} \sqrt{\frac{\theta}{R}},
\end{equation}
where $U_e$, $\theta$ and $R$ are the freestream velocity at the edge of the boundary layer, local momentum thickness, and the radius of curvature of the mean flow respectively, exceeds 0.3~\citep{Gortler54, Smith55}. The most amplified wavelength in the spanwise direction is found to be $\lambda_T$, corresponding to  ${U_e\lambda_T}/{\nu} \sqrt{{\lambda_T}/{R}} \approx 220-270$~\citep{Smith55,FloryanS82,LuchiniB98}, or  $\lambda_T \approx \delta-2\delta$~\citep{SmitsD06}.  However, the above scaling laws are for laminar flows and their applicability to turbulent flows is unclear. \cite{Tani62} proposed that the criterion for the onset of the G\"{o}rtler instability is valid for turbulent flows provided the molecular viscosity is replaced by the eddy viscosity. It has also been suggested that $\delta/R\ge 0.01$ is the applicable criterion instigating G\"{o}rtler instability for the case of turbulent boundary layers~\citep{HoffmannMB85,Floryan91}.

\begin{figure}
\centering
  \includegraphics[width=0.8\textwidth]{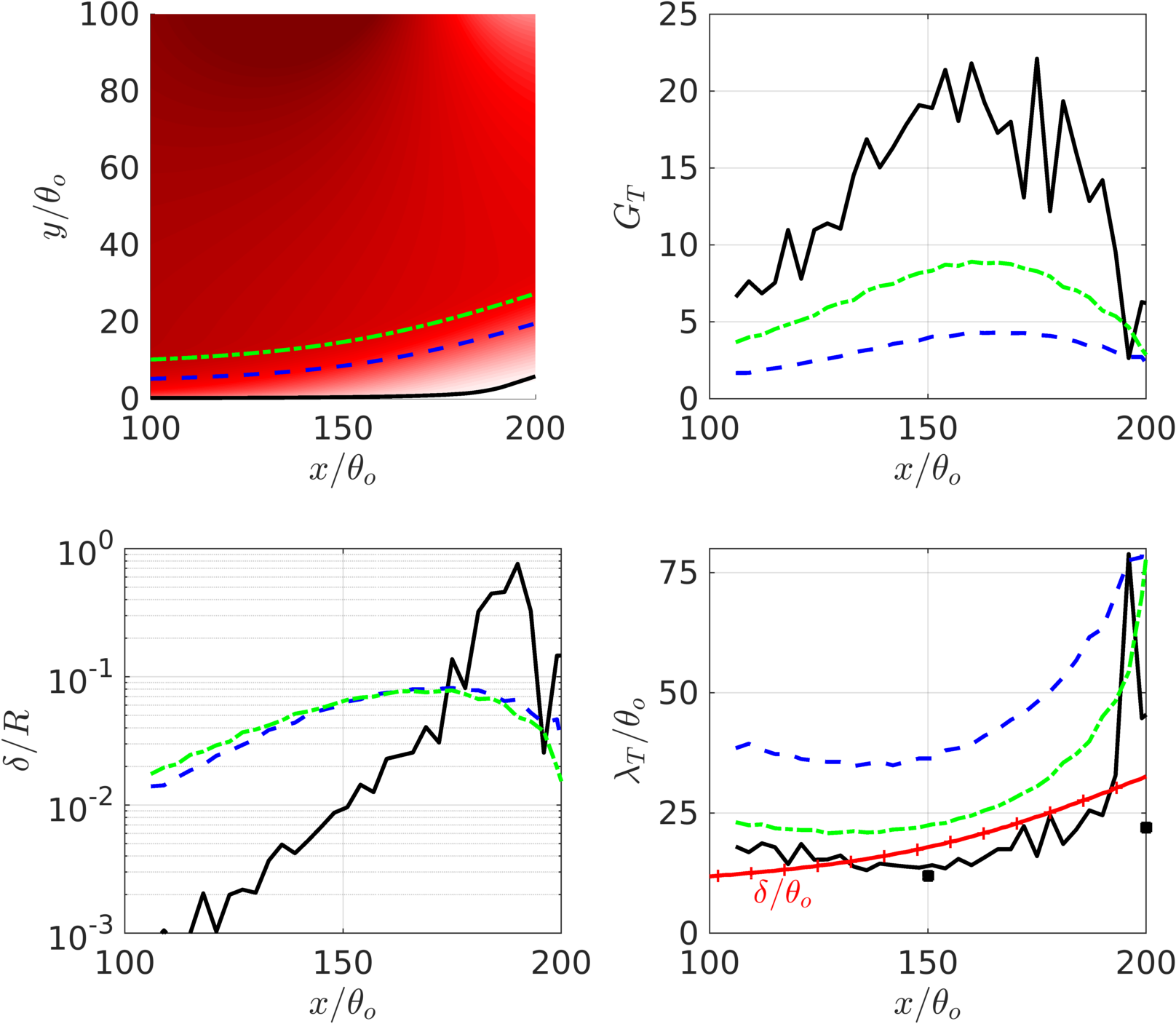}
  \caption{(a) Selected streamlines (solid, dashed and dash-dotted respectively) for G\"{o}rtler instability analysis; (b) G\"{o}rtler number; (c) $\delta/R$; (d) most amplified spanwise wavelength predicted by ${U_e\lambda_T}/{\nu_\text{tot}} \sqrt{{\lambda_T}/{R}} = 250$. Solid with cross is $\delta/\theta_o$ for reference. Black markers are the length scale in the separated shear layer measured by the streamwise fluctuating velocity (Fig. \ref{fig:Rii}).}
  \label{fig:Gt}
\end{figure}
To examine these proposed criteria for the present flow, \revb{we calculate the effective turbulent eddy viscosity as~\citep{SpalartStrelets00} 
\begin{equation}
\nu_{t,\text{eff}} = -\frac{\overline{u^\prime_i u_j^\prime} S_{ij}}{2S_{ij}S_{ij}}
\end{equation}
and use $\nu_{tot} = \nu + \nu_{t,\text{eff}}$} to estimate the G\"{o}rtler number. The analysis is applied in the region of mean streamline concavity shown in Fig. \ref{fig:Gt}. It can be seen that when the boundary layer separates from the wall, the G\"{o}rtler number ranges from about 5 to 21 in the near-wall region and  $\delta/R$ ranges from 0.08 to 0.5. These values are well above the threshold proposed in previous studies, indicating the viability of the occurrence of G\"{o}rtler instability. The most amplified spanwise wavelength (see Fig. \ref{fig:Gt} (d)) is predicted to range from about 25 $\theta_o$ to 50 $\theta_o$ at $x=180\,\theta_o$, which is in the range of 1 to 2 $\delta$. The low-frequency DMD mode exhibits about four pairs of counter-rotating streamwise vortices in the spanwise direction near $x=\,200\theta_o$ (Fig. \ref{fig:udmd_fl} (a)). This corresponds to a wavelength of $0.9\,\delta$, consistent with the predicted range of the most amplified wavelength of possible G\"{o}rtler instability modes.  

It is noted that downstream of $x=200\,\theta_o$ the mean streamlines become convex in shape but the sign of the velocity gradient remain unchanged except for the very near wall region inside the separation bubble. Thus, while the conditions required for the G\"{o}rtler instability no longer exist in this region,  G\"{o}rtler vortices that were generated upstream of the region continue to develop and grow on this region. As observed from the structures of the low-frequency DMD modes, the scale of the elongated streamwise structures grows in the streamwise direction and the structures break down around $x=400\,\theta_o$. 

\subsection{Breakdown of G\"{o}rtler Vortices and Turbulence Structures}
In this section we examine the elongated streamwise structures in the TSB and their possible relationship to structures in the incoming turbulent boundary layer upstream of the separation bubble. The prominence of  large-scale streamwise structures in zero-pressure-gradient TBLs at high Reynolds numbers is well known ~\citep{AdrianMT00,GanapathisubramaniLM03,HutchinsM07}, and these structures are amplified in the presence of APG ~\citep{SkoteHH98,LeeS09,HarunMMM13}. In studies on flow separation that employ suction-and-blowing~\citep{AbeMMS12,Abe17}, \revb{large-scale structures identified by the streamwise velocity fluctuations were also observed but were attributed to the lifted and energized outer layer structures after flow separation}. Our results of the DMD analysis, on the other hand, indicate that the streamwise elongated structures are mostly amplified when the flow separates and streamline curvature becomes significant. This however, does not necessarily imply there is no link between the low-speed regions (``streaks") in the incoming boundary layer and the large-scale streamwise oriented strictures we have identified as G\"{o}rtler vortices. The low-speed regions could serve as the perturbations for the initial development of the G\"{o}rtler vortices. Similar to ~\cite{PriebeTRM16}, we identified a time-delayed correlation between the appearance of low-speed region in the incoming boundary layer and the occurrence of the G\"{o}rtler vortices (not shown). This could be the reason why the occurrence of the elongated structures in the spanwise direction is highly unsteady, instead of being at fixed locations when the G\"{o}rtler instability is triggered by prescribed upstream disturbances~(\cite{FloryanS82} among others).  

\begin{figure}
\centering
  \includegraphics[width=0.95\textwidth]{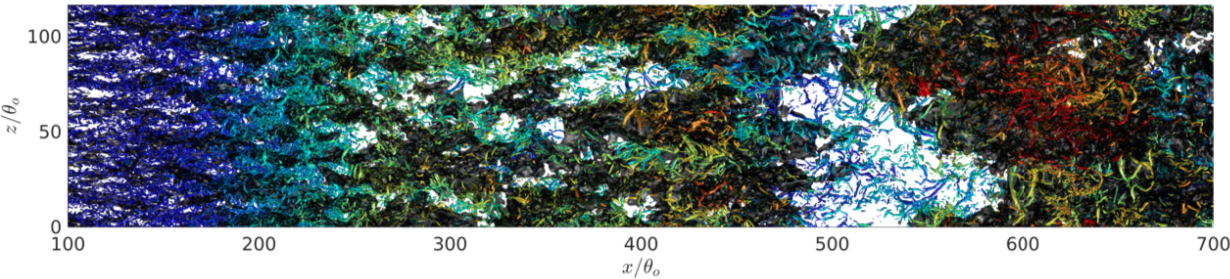}
  \caption{Topview of the instantaneous vortical structures showed by the isosurfaces of the secondary invariant of the velocity-gradient tensor $Q$, colored by the distance from the wall. The dark-gray isosurfaces are $u^\prime=-0.1U_o$. Isometric view and colormap refer to Fig. \ref{fig:Qiso} (b).}
  \label{fig:Q}
\end{figure}
Figure \ref{fig:Q} shows the top view of the turbulent structures exhibited in Fig. \ref{fig:Qiso}. The streamwise alignment of the structures ($x<450\,\theta_o$) is quite evident \revb{and the scale of the low-speed region increases as the flow travels downstream (\textit{e.g.}, the dark gray isosurface of $u^\prime=-0.1U_o$ at $x/\theta_0 = 300$ is much thicker than the ones at $x/\theta_0 = 100$ and 200 which appear to be thin rod-like regions.)}. The growth of G\"{o}rtler vortices in a laminar flow over a concave wall has been examined in previous studies~\citep{BippesG72,SwearingenB83,SmithWalker89,LiM95} and one observation is that the continuous ejection of fluid from the low-speed region by the counter-rotating vortices creates a marked retardation of flow in the upwash region and an APG in the streamwise direction, which has the dual effect of both forcing the vortices away from the wall and also increasing the spanwise distance between the vortices~\citep{BippesG72,SmithWalker89}. The behaviour of the streamwise structures shown in our simulation is in line with this observation:  

\begin{figure}
\centering
  \includegraphics[width=0.8\textwidth]{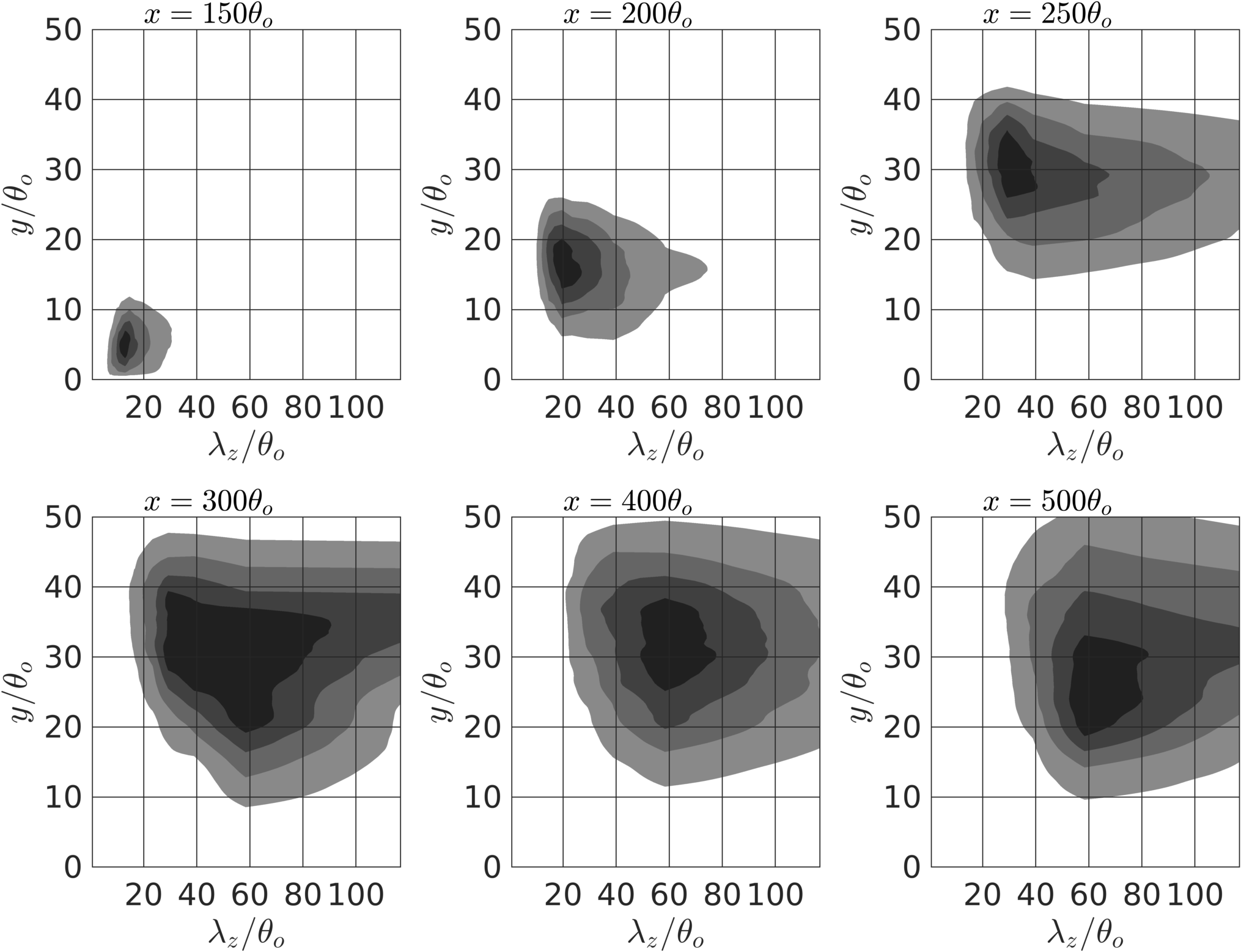}
  \caption{Pre-multiplied spanwise energy spectrum of the streamwise fluctuating velocity, examined at several streamwise locations. Each map is normalized by its maximum and contours are showed at level 0.8, 0.6, 0.4 and 0.2, from dark to light respectively. }
  \label{fig:Rii}
\end{figure}
The scale of the streamwise structures in the spanwise direction is examined using the pre-multiplied energy spectrum of $u^\prime$ and figure \ref{fig:Rii} shows the spectrum at several streamwise locations. The growth of the spanwise lengthscale is clearly seen.
The spanwise lengthscale of these structures is about 20 $\theta_o$ to 30 $\theta_o$ in the 
upstream side of the separation bubble, which agrees reasonably well with the most unstable wavelength predicted by the G\"{o}rtler instability (refer to Fig. \ref{fig:Gt} d). From $x=300\,\theta_o$ the wavelength changes to $60\,\theta$, consistent with the observation of the low-frequency DMD mode which shows that the structures merge after the crest of the separation bubble.

In previous studies, it has been proposed that G\"{o}rtler vortices may break down due to: a) Tollmien-Schlichting waves; b) non-linear development of G\"{o}rtler vortices and energy cascades down from the fundamental into the higher harmonics and the mean flow; c) secondary instabilities that generate a retarded region in the upstream flow region between G\"{o}rtler vortices and the formation of transverse vortices~\citep{BippesG72,Hall82,SmithWalker89}.
Linear stability theory has revealed two types of secondary instabilities that help break down the G\"{o}rtler vortices in laminar flows: a sinuous (odd) mode that perturbs the G\"{o}rtler vortices in a wavy manner, and a varicose (even) mode that breaks up G\"{o}rtler vortices into a series of ``knotty" structures (see \cite{SwearingenB83} and \cite{LiM95} among others). None of these processes is clearly evident in the single DMD mode reconstruction, and the structures in the flow velocity seem to exhibit both streamwise waviness and break up. Which mode (if any) dominates the break up of vortices in this flow is unclear from the current simulations; 
\revc{d) the vortex merging/pairing process. As discussed regarding the frequency change (Fig. \ref{fig:sptialtemporal_PSD}) and the vortex development (Fig. \ref{fig:2pcoef}, \ref{fig:contourQ}), it is possible that the break-up of the G\"{o}rtler vortices is associated with the merging/pairing of the spanwise roller vortices. 
The DMD mode at the low-frequency does exhibit a spanwise staggered signature (Fig. \ref{fig:udmd_fl} (b)) which provides further support for this possibility. However, this correlation does not necessarily imply causation.
%The DMD analysis here shows a correlation between this mode and the G\"{o}rtler vortices, but does not imply causation. The DMD mode at the low-frequency does exhibit spanwise staggered behaviour (Fig. \ref{fig:udmd_fl} (b)) and this may indicate that the merger/pairing could be modulated by the G\"{o}rtler vortices.
}

\section{Conclusions}
Direct numerical simulations of a turbulent boundary layer over a flat plate with induced separation are performed with the aim of investigating the spatio-temporal dynamics of turbulent separation bubbles. The separation is induced by employing a transpiration boundary condition on the top boundary of the computational domain. Two different transpiration velocity profiles are employed: one with suction followed by blowing and the other with suction only. Compared with the TSB created by the suction-blowing profile, the suction-only case exhibits a pressure gradient and Reynolds stress distribution which is in much better qualitative agreement with separated flows over airfoil and in diffusers. 

Focusing on the suction-only case, we show that the high and mid-frequency motion is well characterized by flow physics that corresponds to that observed in a plane turbulent mixing layer. In particular, these modes are associated with the formation and shedding of spanwise oriented rollers, as also confirmed by DMD analysis. The characteristic frequency of these modes does not scale with $U/L_{sep}$ because in both the TSB-SB and TSB-SO cases this mode occurs at a similar frequency, while the length of the mean separation bubble differs by a factor of about 2.4.

\revc{The suction-only case exhibits a low-frequency unsteadiness (i.e., breathing/flapping) in the separation bubble at a frequency which is two and a half times smaller than the dominant high-frequency. A similar low-frequency motion is not observed in the  suction-blowing case. The low-frequency motion observed here appears as a spatio-temporal variation of the separation region: the separation bubble opens at regular intervals corresponding to this low frequency and releases a large-scale conglomeration of vortices that convects downstream and is associated with low-speed and even reversed flow. 
DMD analysis of the flow, however, shows that the topological signature of this mode does not simply correspond to the merging of spanwise rollers. 
The single DMD mode at the low-frequency exhibits a topology that is dominated by highly elongated streamwise structures that extend from nearly the point of separation to downstream of the mean reattachment point. Analysis of the data shows that the streamline curvature does exceed the threshold for G\"{o}rtler instability and the properties of the elongated streanwise structures, such as spanwise wavelength, agree with the values reported in the literature. Their periodic breakdown, possibly a secondary instability, could cause the observed low frequency mode or modulate the vortex merger.}
\reva{The absence of the low-frequency motion in the TSB-SB case may be due to the forced closure (reattachment) of the separation bubble which does not allow natural instabilities, especially those with large time and length scales to grow.}

The simulations however do not indicate the mechanism for the breakdown of the  elongated streamwise structures and the subsequent formation of the large scale eddies that are released from the separation bubble. The scale separation between the low and high frequencies in the current study is limited due to the relatively low Reynolds numbers. 
%------------
\reva{The K-H instability is inviscid in nature and the characteristic vortex shedding frequency scales with the outer scales. Therefore, the high- and medium-frequency motions are not expected to change much with the Reynolds number. In the simulation by \cite{Abe17}, the peaks in the spectra of  wall pressure do not change much between $Re_\theta=$ 300 and 900 (refer to Fig. 21 in \cite{Abe17}, given the minor change in $L_{sep}$.}
%------------
Ongoing studies are focused on higher Reynolds numbers where a larger scale-separation will enable a clearer distinction between the dominant spatio-temporal scales in the flow.  
Lastly, the effects of streamline curvature cannot be dismissed in the separation bubble but the curvature can be altered by changing the magnitude of the APG. A very mild APG may cause small streamline curvature and prevent the amplification of the G\"{o}rlter vortices. Some of these issues are presently under study.
 
\section*{Acknowledgments}
The authors would like to thank Drs. L. Cattafesta (Florida State University), C. Rowley (Princeton University) and D. Gayme (Johns Hopkins University) for fruitful discussions.   The authors also acknowledge the support from AFOSR Grant FA9550-17-1-0084, monitored by Dr. Gregg Abate. The simulations were performed at the Texas Advanced Computing Center (TACC) Stampede-2 cluster. 

\section*{Appendix: Dynamic Mode Decomposition}
%---------------------------
The DMD analysis and field reconstruction are performed by the following procedure:
\begin{enumerate}[align=left,leftmargin=3em,itemindent=0pt,labelsep=0pt,labelwidth=2em]
\item Collect snapshots of data ($\boldsymbol{\phi} = [\vec{\phi}_1,\vec{\phi}_2,...,\vec{\phi}_{N}]$, where each column of $\boldsymbol{\phi}$ consists of vector of observables (reshaped to 1D if sampled in 2D/3D) at $t_i$ and $N$ the total number of snapshots) from the simulation at several time instances that are equally spaced in time. $\vec{\phi}$ might correspond for instance to the velocity components or pressure for a set of grid point in a region of the flow. Each column $\vec{\phi}$ can include several observables. If the observables are sampled at $M$ spatial locations and a total of $P$ observables are used, the size of $\boldsymbol{\phi}$ is $[M\times P,\,N]$.
\item Arrange the data $\boldsymbol{\phi}$ into matrices \\
\centerline{ $\boldsymbol{X}\equiv [\vec{\phi}_1, \vec{\phi}_2, ..., \vec{\phi}_{N-1}]$, $\boldsymbol{Y}\equiv [\vec{\phi}_2, \vec{\phi}_3, ..., \vec{\phi}_{N}]$}
\item Dimensionality reduction: compute the (reduced) SVD of $\boldsymbol{X}$, i.e.: \\ 
\centerline{$\boldsymbol{X}^\prime = \boldsymbol{U}\boldsymbol{\Sigma} \boldsymbol{V}^{T}$}
where $\boldsymbol{U}$ is of size $[M \times P, r]$, $\boldsymbol{\Sigma}$ is $r \times r$, and $\boldsymbol{V}$ is $[N, r]$, where $r$ is the reduced rank of $\boldsymbol{X}^\prime$, or the total number of the conjugated modes. %
\item Define the matrix (size $[r, r]$)\\
\centerline{$\boldsymbol{\tilde{A}}\equiv \boldsymbol{U}^{T}\boldsymbol{Y}\boldsymbol{V}\boldsymbol{\Sigma}^{-1}$}
\item Compute eigenvalues ($\lambda$) and eigenvectors ($\vec{w}$) of $\boldsymbol{\tilde{A}}$, writing\\
\centerline{$\boldsymbol{\tilde{A}} \vec{w} = \lambda \vec{w}$}.
\item The DMD mode corresponding to the DMD eigenvalue $\lambda$ is then given by\\
\centerline{$\boldsymbol{\psi} \equiv \boldsymbol{Y}\boldsymbol{V}\boldsymbol{\Sigma}^{-1}\vec{w}$}.
In this study, the modes are normalized by their initial amplitudes in $\vec{\phi}_1$.
\item Calculate the (initial) amplitude of each mode, \textit{i.e.}, contribution of each mode to the first snapshot. \\
\centerline{$\vec{b}=\boldsymbol{\psi}^{-1}\vec{\phi}_1$}.
\item Obtain the frequency of each mode\\
\centerline{$f={\Im}(\text{ln}(\lambda))/(2\pi\Delta t)$}
or, {$f=\text{tan}^{-1} \left[ {\Im}(\boldsymbol{\psi})/{\Re}(\boldsymbol{\psi})\right]/(2\pi\Delta t)$}. 
\item Define $\omega = \text{ln}(\lambda)/\Delta t$, reconstruct the data series by \\
\centerline{$\boldsymbol{\phi}_{dmd}(t) \equiv \boldsymbol{\psi} \text{exp}\left(\vec{\omega} t\right) \vec{b}$}
where in the right-hand-side, $\boldsymbol{\psi}$ gives the shape of the mode, $\vec{b}$ determines the magnitude, and the exponential term represents decay/growth. 
\end{enumerate}

%\nocite{*}
%\clearpage
\bibliographystyle{./jfm}
%\bibliography{/Users/ugo/Dropbox/Research/Biblio}
\bibliography{myref}

\end{document}